\documentclass[paper,11pt]{article}

\makeatletter
\@addtoreset{equation}{section}
\makeatother

\usepackage{epsfig,amsfonts}
\usepackage{graphicx}
\usepackage{caption}
\usepackage{subcaption}
\usepackage{amsmath}
\usepackage{amsthm,amssymb}
\usepackage{mathrsfs}
\usepackage{hhline}
\usepackage{upgreek}
\usepackage[ruled,vlined]{algorithm2e}
\usepackage{hyperref}
\usepackage[dvipsnames]{xcolor}
\usepackage{musicography}
\usepackage[normalem]{ulem}
\definecolor{purple_nice}{rgb}{0.4,0.2,0.7}
\definecolor{fuel_blue}{RGB}{42,162,185}
\definecolor{YInMn_blue}{RGB}{46, 80, 144}
\definecolor{ultramarine}{RGB}{63, 0, 255}
\definecolor{KLEIN_blue}{rgb}{0, 0.18, 0.65}
\hypersetup{
    colorlinks=true,
    linkcolor=YInMn_blue,
    citecolor=ultramarine,
    filecolor=fuel_blue,
    urlcolor=KLEIN_blue,
}

\renewcommand{\Re}{\operatorname{Re}}
\renewcommand{\Im}{\operatorname{Im}}
\usepackage{enumitem}

\def\be{\begin{equation}}
\def\ee{\end{equation}}
\def\bea{\begin{eqnarray}}
\def\eea{\end{eqnarray}}

\def\XXint#1#2#3{{\setbox0=\hbox{$#1{#2#3}{\int}$}
    \vcenter{\hbox{$#2#3$}}\kern-.5\wd0}}

\def\Tb{\bar T}

\begin{document}

\begin{titlepage}

\title{\vspace{-3cm}{\huge On Factorizable S-matrices, Generalized TTbar, \break and the Hagedorn Transition}}
\author{Giancarlo Camilo$^{1,2\musDoubleFlat}$, Thiago Fleury$^{1\musFlat}$, M\'{a}t\'{e} Lencs\'{e}s$^{3\musNatural}$,\\[0.1cm] Stefano Negro$^{4\musSharp}$ and Alexander Zamolodchikov$^{5,6\musDoubleSharp}$\\[0.3cm]}
\date{\footnotesize{$^1$ 
International Institute of Physics, Federal University of Rio Grande do Norte\\ Campus Universit\'ario, Lagoa Nova, 59078-970, Natal, RN, Brazil\\[0.1cm]$^2$ Quantum Research Centre, Technology Innovation Institute, Abu Dhabi, UAE\\[0.1cm]$^3$ BME-MTA Statistical Field Theory `Lend\"ulet’ Research Group
\\Department of Theoretical Physics, Budapest University of Technology and Economics 
 \\1111 Budapest, Budafoki {\'u}t 8, Hungary\\[0.1cm]$^4$ Center for Cosmology and Particle Physics, New York University, New York,\\ NY 10003, U.S.A.
\\[0.1cm]$^5$ C.N. Yang Institute for Theoretical Physics, State University of New York, Stony Brook,\\ NY 11794-3840, USA.\\[0.1cm]$^6$ Kharkevich Institute for Information Transmission Problems, Moscow, Russia\\[0.3cm]$^{\musDoubleFlat}$\texttt{\href{mailto:gcamilo@iip.ufrn.br}{gian.fis@gmail.com},} $^{\musFlat}$\texttt{\href{mailto:tsi.fleury@gmail.com}{tsi.fleury@gmail.com},} $^{\musNatural}$\texttt{\href{mailto:mate.lencses@gmail.com}{mate.lencses@gmail.com},}\\ $^{\musSharp}$\texttt{\href{mailto:stefano.negro@nyu.edu}{stefano.negro@nyu.edu},} $^{\musDoubleSharp}$\texttt{\href{mailto:alexander.zamolodchikov@stonybrook.edu}{alexander.zamolodchikov@stonybrook.edu},}\\}}
\maketitle

\begin{abstract}
We study solutions of the Thermodynamic Bethe Ansatz equations for relativistic theories defined
by the factorizable $S$-matrix of an integrable QFT deformed by CDD factors. 
Such $S$-matrices appear under generalized TTbar deformations of integrable QFT by special irrelevant operators. The TBA equations, of course, determine the ground state energy $E(R)$ of the finite-size system, with the spatial coordinate compactified on a circle of circumference $R$. We limit attention to theories involving just one kind of stable particles, and consider deformations of the trivial (free fermion or boson) $S$-matrix by CDD factors with two elementary poles and regular high energy asymptotics -- the ``2CDD model". 
We find that for all values of the parameters (positions of the CDD poles) the TBA equations exhibit two real solutions at $R$ greater than a certain parameter-dependent value $R_*$, which we refer to as the primary and secondary branches. The primary branch is identified with the standard iterative solution, while the secondary one is unstable against iterations and needs to be accessed through an alternative numerical method known as pseudo-arc-length continuation.
The two branches merge at the ``turning point" $R_*$ (a square-root branching point). The singularity signals a Hagedorn behavior of the density of high energy states of the deformed theories, a feature incompatible with the Wilsonian notion of a local QFT originating from a UV fixed point, but typical for string theories. 
This behavior of
$E(R)$ is qualitatively the same as the one for standard TTbar deformations of local QFT.  

\end{abstract}
\end{titlepage}
\newpage

\tableofcontents

\newpage
\section{Introduction}\label{sec:intro}
The so-called TTbar deformations \cite{Smirnov:2016lqw,Cavaglia:2016oda} of two-dimensional quantum field theories (QFTs) has brought about a renewed interest to UV properties of Renormalization Group (RG) flows generated by higher dimensional (a.k.a. ``irrelevant'') operators. The TTbar deformation is defined as the one-parameter family of formal ``actions'' $\mathcal{A}_\alpha$, determined by the flow
\begin{eqnarray}\label{Aalpha}
\frac{d}{d\alpha}\mathcal{A}_\alpha = \int\,(T\Tb)_\alpha (x)\,d^2 x\;,
\end{eqnarray}
where $(T\Tb)_\alpha (x)$ is a special composite operator built from the components of the energy-momentum tensor associated with the theory $\mathcal{A}_\alpha$ \cite{Zamolodchikov:2004ce}. The deformation \eqref{Aalpha} has a number of notable properties. The theory $\mathcal{A}_\alpha$ is ``solvable'', in the sense that certain characteristics can be found exactly in terms of the corresponding ones in the undeformed theory $\mathcal{A}_{\alpha=0}$. This is remarkable, because the deformation operator
$(T\Tb)_\alpha$ has exact dimension $4$, meaning the perturbation in \eqref{Aalpha} is ``irrelevant'' in the RG sense. Normally, such deformations are expected to break the short-distance structure of the quantum field theory, generally rendering the theory UV incomplete, and possibly violating causality at short scales. The abnormal UV properties of the theory $\mathcal{A}_\alpha$ are manifest already in the short-scale behavior of its finite-size ground-state energy. If the spatial coordinate of the 2D space-time is compactified on a circle of circumference $R$, its ground-state energy $E_\alpha(R)$ is determined exactly, via the equation \cite{Smirnov:2016lqw,Cavaglia:2016oda}
\begin{eqnarray}\label{burgers2}
E_\alpha(R)=E_0(R-\alpha E_\alpha(R))\;,
\end{eqnarray}
in terms of the ground state energy $E_0(R)$ of the undeformed theory, at $\alpha=0$. The equation \eqref{burgers2} shows that, depending on the sign of the deformation parameter $\alpha$, the ground state energy either develops a square root singularity at some $R_*\sim 1/\sqrt{|\alpha|}$, or has no short-distance singularity at all.
Neither of these types of behavior is compatible with the idea of QFT as the RG flow stemming out of a UV fixed point. The theory defined by \eqref{Aalpha} therefore is not a local QFT in the Wilsonian sense \cite{Wilson:1973jj}. Moreover, at negative $\alpha$, the singularity at finite $R$ signals a very fast growth of the density of states at high energies, a common hallmark of string theories, leading to the Hagedorn transition \cite{Polchinski:1998rq}. The behavior of $E_\alpha(R)$ at positive $\alpha$ is possibly even more puzzling, as it suggests a finite number of states per unit volume, an unlikely feature if one thinks of a QFT as a system of continuously many interacting degrees of freedom, unless quantum gravity is involved\footnote{A relation of the TTbar deformation to the Jackiw-Teitelboim gravity was indeed proposed in \cite{Dubovsky:2017cnj,Dubovsky:2018bmo}.}. Therefore, the deformed theory determined by \eqref{Aalpha} cannot be considered a conventional UV complete local QFT. At the same time, however, the TTbar deformation has a number of robust features which makes one reluctant to simply dismiss it as ``pathological''. It is instead tempting to think that the deformation \eqref{Aalpha} exemplifies some meaningful extension of the notion of local QFT. In particular, an interesting interpretation of the theory $\mathcal{A}_\alpha$ in terms of its gravitational dual was proposed in \cite{McGough:2016lol}, where a relation to the state of the bulk gravity in the dual theory was suggested.
Several questions about 2D physics of the deformed theory need to be elucidated in order to put such suggestions on a solid ground. For example, does the deformation preserve any part of the local structure of QFT? Notice how the very definition \eqref{Aalpha} depends on the notion of the energy-momentum tensor, conventionally a part of such a local structure. Another important question concerns the macro-causality in 2D space-time. While the deformation \eqref{Aalpha} with positive $\alpha$ is suspected to display super-luminal propagation \cite{Dubovsky:2017cnj,McGough:2016lol}, the case of negative $\alpha$ is most likely free from this problem. We will not dwell on this question, as it is the negative-$\alpha$ deformation which will be of interest to the present discussion. In any case, we believe it is important to understand the physical origin of the above abnormal short-distance properties.

Another exact result about the theory $\mathcal{A}_\alpha$ concerns the deformation of its $S$-matrix, whose elements differ from
the corresponding undeformed ones by a universal phase factor, available in closed form \cite{Dubovsky:2017cnj}. In particular, the $2 \to 2$ elastic scattering amplitude has the form
\begin{eqnarray}\label{sdef}
S_\alpha(\theta) = S_0(\theta)\,\exp\left(-i \alpha M^2 \sinh\theta\right)\;,
\end{eqnarray}
where $S_0(\theta)=S_{\alpha=0}(\theta)$ is the $2\to 2$ scattering amplitude of the undeformed theory. Here $\theta=\theta_1 - \theta_2$ is the difference between rapidities of the two particles involved -- assumed for simplicity to be identical -- and $M$ denotes their mass; in what follows we set the units so that $M=1$. A notable feature of the additional phases acquired under the deformation is their abnormally fast high-energy growth, which is evident already in the form \eqref{sdef}\footnote{A similar behavior of the scattering phase was previously found in non-commutative field theories \cite{Douglas:2001ba}.}. The scattering phase in \eqref{sdef}
determines the density of two-particle states, suppressing it when $\alpha>0$ but greatly enhancing it at negative $\alpha$. In the latter case, one might be led to believe that the Hagedorn behavior is directly related to this rapid growth of the $2\to 2$ scattering phase. One of the results of the present work is to show that the situation is more subtle: the growth of the two-particle scattering phase in \eqref{sdef} is not a necessary condition for the formation of the singularity of the finite-size energy at finite real $R$. We will study certain generalizations of the TTbar deformation which can be defined whenever the original QFT is integrable \cite{Smirnov:2016lqw}. In most of such deformations, the scattering phases present a less exotic high-energy behavior  -- i.e., they have finite limit at $\theta\to\infty$ -- while, at the same time, the overall density of states grows nonetheless exponentially with the energy, leading to the Hagedorn singularity.

The generalizations of the TTbar deformations we will be interested in are based on the integrability of the original QFT. This assumes that the theory possesses infinitely many conserved local currents of higher Lorentz spins $s+1$, with $s$ taking values in the set $\{s\}$ of odd natural numbers: $s=1,3,5,7,...$\footnote{Generally, the set of spins $\{s\}$ of local Integrals of motion may be different in different integrable theories. Here we assume, again for simplicity, the most common situation -- represented e.g. by sinh-Gordon or sigma models -- where $\{s\}$ involves all odd natural numbers. In different models the CDD factor discussed below may be constrained by additional conditions, which however do not change the overall conclusions below.}. The deforming operators $T\Tb^{(s)}(x)$ are constructed from these currents in the exact same way as the operator $T\Tb (x)$ is built from the energy-momentum tensor, see \cite{Smirnov:2016lqw} for details. It can be then shown that the theory deformed by adding such operators retains its integrability, preserving the same set of conserved local currents. Therefore the deformations of an Integrable QFT (IQFT) by the operators $T\Tb^{(s)}$ generate an infinite-dimensional family of flows generalizing \eqref{Aalpha},
\begin{eqnarray}\label{Aalphas}
\frac{\partial\mathcal{A}_{\{\alpha\}}}{\partial\alpha_s} = \int\,T\Tb^{(s)}_{\{\alpha\}}(x)\,d^2 x\;.
\end{eqnarray}
Here $\{\alpha\}$ denotes the infinite set of the deformation parameters $\{\alpha\}:=\{\alpha_s\}$, and the subscript $\{\alpha\}$ under the operator $T\Tb^{(s)}(x)$ is added to emphasize that it is constructed in terms of the conserved currents of the deformed theory $\mathcal{A}_{\{\alpha\}}$. In what follows we refer to \eqref{Aalphas} as the \emph{generalized TTbar flow}\footnote{In \cite{Conti:2019dxg}, a different family of  generalizations of the TTbar flow, in which the deforming operators $T\Tb_{s}$ are asymmetrically constructed from the energy-momentum tensor and a higher-conserved current, was explored.}. 
For integrable theories, the infinite-parameter flow \eqref{Aalphas} generalizes the one-parameter deformation \eqref{Aalpha}. The latter corresponds to the special case $\alpha_s =0$ for $s>1$, and 
$\alpha_1=\alpha$. To distinguish them, below we often refer to \eqref{Aalpha} as the "TTbar proper", or simply TTbar, reserving the term "Generalized TTbar" to the generic deformation \eqref{Aalphas}.
It was argued that the deformation \eqref{Aalphas} leads to the following deformation of the elastic two-particle $S$-matrix 

\begin{eqnarray}\label{sdefs}
S_{\{\alpha\}}(\theta)=S_{\{0\}}(\theta)\,\Phi_{\{\alpha\}}(\theta)\,,
\qquad \Phi_{\{\alpha\}}(\theta)=\exp\left\{-i \,\sum_{s\in2\mathbb{Z}+1}\,\alpha_s\,\sinh\left(s\,\theta\right)\right\}\;,
\end{eqnarray}
with the same notations as in \eqref{sdef} and \eqref{Aalphas}\footnote{The parameters $\alpha_s$ in \eqref{sdef} coincide with the flow parameters defined in \eqref{Aalphas} provided a specific normalization of the fields $T\Tb^{(s)}_{\{\alpha\}}(x)$ is chosen, otherwise the terms in the sum in \eqref{sdef}
would have additional normalization-dependent numerical coefficients. The form \eqref{sdef} was explicitly derived in \cite{Smirnov:2016lqw} for the deformed sine-Gordon model, to leading order in the deformation parameters. However, this form of the $S$-matrix deformation under the flow \eqref{Aalphas} can be proven in the general case, using the methods of \cite{Cardy:2018sdv} or the approach developed in \cite{Kruthoff:2020hsi}. We will elaborate this point elsewhere.}. The phase factor $\Phi_{\{\alpha\}}(\theta)$ is known with the name of \emph{CDD factor} \cite{Castillejo:1955ed}. Generally, it is an energy-dependent phase factor $\Phi(\theta)$ which can be added to the $2\to 2$ scattering amplitude without violating the analyticity, unitarity and crossing symmetry conditions. The unitarity and crossing demand that $\Phi(\theta)$ satisfies the functional relations
\begin{eqnarray}\label{cdddef}
\Phi(\theta)\Phi(-\theta)=1\,, \qquad \Phi(\theta)=\Phi(i\pi-\theta)\;,
\end{eqnarray}
which $\Phi_{\{\alpha\}}(\theta)$ in \eqref{sdefs} obviously does term by term in the sum over $s$. 
Moreover, it is easy to see that (once the overall sign ambiguity is ignored) any solution of \eqref{cdddef} can be represented by the form \eqref{sdefs}, with the series in the exponential converging in some vicinity of the point $\theta=0$.
However, the series does not need to converge at all $\theta$. The $S$-matrix analyticity forces $\Phi(\theta)$ to be a meromorphic function of $\theta$, with the locations of the poles constrained by the condition of macro-causality (more on this momentarily). Therefore, for \eqref{sdefs} to represent a physically sensible $S$-matrix, the sum over $s$ is allowed to have a finite domain of convergence, while its analytic continuation must admit the representation
\begin{eqnarray}\label{cddg}
\Phi_{\{\alpha\}}(\theta) = \Phi_{\text{pole}}(\theta) \,\Phi_{\text{entire}}(\theta)\;,
\end{eqnarray}
where the first factor absorbs all the poles located at finite $\theta$, whose number $N$ is in general arbitrary (possibly infinite),
\begin{eqnarray}\label{phipole}
\Phi_{\text{pole}}(\theta) = \prod_{p=1}^N  \frac{\sinh\theta_p + \sinh\theta}{\sinh\theta_p - \sinh\theta}\;,
\end{eqnarray}
and
\begin{eqnarray}\label{phientire}
\Phi_{\text{entire}}(\theta) = \exp\left\{-i \,\sum_{s}\,a_s \,\sinh\left(s\,\theta\right)\right\}\;.
\end{eqnarray}
In this last factor, the series in the exponential is assumed to converge at all $\theta$, so that $\Phi_{\text{entire}}(\theta)$ represents an entire function of $\theta$. Macro-causality restricts possible positions of the poles $\theta_p$ to either the imaginary axis $\Re \theta_p=0$, or to the strips $\Im \theta_p \in [-\pi,0] \ \text{mod} \ 2\pi$ since, in virtue of \eqref{cdddef}, $\Phi(\theta)$ is a periodic function, $\Phi(2\pi i+\theta)=\Phi(\theta)$. 
Let us stress here that the representation (\ref{cddg}--\ref{phientire}) of the generic CDD factor $\Phi_{\{\alpha\}}(\theta)$ differs from the one given in \eqref{sdefs} only in the parameterization: any factor (\ref{cddg}--\ref{phientire}) can be written in the form \eqref{sdefs}, with the parameters $\alpha_s$ expressed in terms of $a_s$ and $\theta_p$, and conversely any factor $\Phi_{\{\alpha\}}(\theta)$ defined in \eqref{sdefs}, being
analytically continued to the whole $\theta$-plane, can be written in the form \eqref{cddg}.

In the present work we focus our attention on the class of $S$-matrices \eqref{sdefs} having CDD factors \eqref{cddg} for which the entire part \eqref{phientire} is absent\footnote{A first analysis of models whose $S$-matrix is deformed by a CDD factor consisting of only of a generic entire part \eqref{phientire} has been performed in \cite{Hernandez-Chifflet:2019sua}.},
\begin{eqnarray}\label{cddn}
\Phi_{\{\alpha\}}(\theta)= \Phi_\text{pole}(\theta)\;,
\end{eqnarray}
and the product in \eqref{phipole} involves finitely many factors, i.e. $N<\infty$. Note that, unlike \eqref{sdef}, such CDD factors have regular limits at $\theta\to\pm \infty$. Therefore, if the undeformed $S$-matrix $S_0(\theta)$ behaves regularly -- presenting no abnormal growth of the scattering phase -- at large $\theta$, so does the deformed $S$-matrix $S_0(\theta)\Phi(\theta)$. We now raise the following question: how does an $S$-matrix deformation such as the one just described affect the short-distance behavior of the theory? Unfortunately, for the general TTbar deformation \eqref{Aalphas} no closed form of the finite-size energy levels similar to \eqref{burgers2} is available with which one could analyze their dependence on the size $R$ of the system. However, having an exact expression for the deformed IQFT $S$-matrix, the finite-size ground-state energy $E(R)$ can be obtained by solving the associated Thermodynamic Bethe Ansatz (TBA) equation \cite{Yang:1968rm,Zamolodchikov:1989cf}.

In general, the form of the TBA equations depends on the particle spectrum of the theory. Here we consider, for simplicity, the case of a factorizable $S$-matrix involving only one kind of particles, having mass $M=1$. In this case the two-particle $S$-matrix consists of a single amplitude $S(\theta)$, which itself satisfies the equations \eqref{cdddef}. Therefore we can limit attention to the functions $S(\theta)$ of the form \eqref{phipole}\footnote{One can think of these as CDD deformations of the free $S$-matrix $S(\theta)=\pm 1$.}. There are two substantially different cases, depending on the sign of $S(0) = \sigma = \pm 1$. Following \cite{Zamolodchikov:1989cf}, we refer to these cases as the ``bosonic TBA'' when $\sigma=+1$ and ``fermionic TBA'' for $\sigma=-1$. Given $S(\theta)$, let $\varphi(\theta)$ be the derivative of the scattering phase,
\begin{eqnarray}\label{varphidef}
\varphi(\theta) = \frac{1}{i}\,\frac{d}{d\theta} \log S(\theta)\;.
\end{eqnarray}
Then the TBA equation takes the form of a non-linear integral equation for a single function $\epsilon(\theta)$, the \emph{pseudo-energy},
\begin{eqnarray}\label{tbas}
\epsilon(\theta)=R\,\cosh\theta - \int\,\varphi(\theta-\theta')\,L(\theta')\,\frac{d\theta'}{2\pi}\;, \qquad
\end{eqnarray}
where
\begin{eqnarray}\label{Ldef}
L(\theta) := -\sigma\,\log\left(1-\sigma\,e^{-\epsilon(\theta)}\right)\;.
\end{eqnarray}
The ground state energy can then be recovered from the pseudo-energy via the following expression 
\begin{eqnarray}\label{etbas}
E(R) = -\,\int_{-\infty}^{\infty} \,\cosh\theta\,L(\theta)\,\frac{d\theta}{2\pi}\;.
\end{eqnarray}

In most cases the TBA equations are impervious to the exact analytic derivation of their solutions but are amenable to numerical approaches. These can yield important insight into high energy, \emph{viz.} short distance, properties of the deformed theories \eqref{Aalphas}. A numerical solution can be obtained, with practically arbitrary accuracy, by numerical integration of \eqref{tbas}. This approach was employed to obtain $E(R)$ in a number of IQFT's with known $S$-matrices, see e.g. \cite{Zamolodchikov:1989cf,Zamolodchikov:1991pc}. Usually, the numerical solution is obtained by iterations, starting from a seed function, conventionally taken to be $\epsilon(\theta)=R\,\cosh\theta$, and successively substituting the result of the previous iteration in the right-hand-side in \eqref{tbas}. We will review this approach in \S \ref{subsec:iterative}. If one considers the $S$-matrix associated with a UV complete local IQFT -- such as a conformal field theory (CFT) perturbed by a relevant operator, the sine-Gordon model, or an integrable sigma-model -- the iterations turn out to converge for all $R>0$, and the resulting ground-state-energy $E(R)$ happens to be analytic at all positive real $R$, developing a Casimir singularity at $R=0$. But how adding a CDD factor to the $S$-matrix will affect the TBA solution? This question was addressed in the early 90's by Al. Zamolodchikov, who has considered the modification of the trivial fermionic $S$-matrix $S(\theta) = -1$ by the simplest possible rational CDD factor, namely \eqref{phipole} with $N=1$. In the resulting theory, the celebrated ``staircase model'' \cite{Zamolodchikov:1991pc}, the iterative solution of the TBA still converges at all positive $R$, producing a ground-state-energy $E(R)$ analytic for $R>0$. He also observed that when adding more general CDD factors the situation changes qualitatively. Typically, the convergence of the iterative solution breaks down at $R$ below a certain critical value $R_*$, and the form of the numerical solution at $R>R_*$, where the iterations converge, strongly indicates the existence of a square-root singularity of $E(R)$ at $R_*$ \cite{Zamolodchikov_unpublished}. A similar observation was made in \cite{Mussardo:1999aj}, where a particular CDD deformation of the trivial bosonic $S$-matrix $S(\theta) = 1$ was studied and the numerical solution of the associated TBA equation was found consistent with the existence of a singularity at finite $R_*>0$. We wish to stress that the presence of the singularity at finite $R_*$ and, moreover, its square-root character, are features very similar to the ones displayed by $E(R)$ in the TTbar deformed QFTs, as shown in Fig \ref{ERplotTTbar} below.

In this work we study a few simple cases of CDD deformed TBA equations, using a refined numerical routine based on the so-called ``pseudo-arc-length continuation'' (PALC) method. This allows one to recover solutions to the TBA equation \eqref{tbas} which are unstable under the standard iterative approach. This method is explained in detail in \S \ref{sec:num_meth}. The object of our attention will be trivial $S$-matrices $S(\theta) = \sigma = \pm 1$ deformed by CDD factors
\eqref{phipole} with $N=1,2$. The case $N=1$ with $\sigma=-1$ corresponds to either the sinh-Gordon or the staircase model, depending on the position of the pole. As mentioned just above, these models do not display any abnormal short-distance behavior and were extensively studied in the literature. The bosonic TBA with $N=1$ was considered in \cite{Mussardo:1999aj} and we will comment on it in \S \ref{sec:results}, along with the $N=2$ case. Of these, we mostly address the fermionic cases, although some results for the bosonic TBA are also presented. We find that for all allowed values of the parameters $\theta_p\, (p=1,2)$ the fermionic TBA equation \eqref{tbas} with sufficiently large $R$ possesses two real solutions, or ``branches'', which merge at some finite $R=R_*$. For $R<R_*$ these branches are likely to continue as a pair of conjugated complex-valued solutions. Of these two real solutions at $R>R_*$, one reproduces the iterative solution of the TBA equations \eqref{tbas}. We will call this solution the ``primary branch'', while referring to the other one as the ``secondary branch''. 
Let us stress here that it is the primary branch which directly corresponds to the deformed theory: $E(R)$ on the primary branch represents the finite-size vacuum energy of the deformed theory (in particular, at $R\to\infty$ the effect of the deformation disappears, as expected); it also gives the specific free energy of the deformed theory at temperature $T=1/R$  (in particular, it is the primary branch solution which correctly sums up the virial expansion associated with the input particle theory). In this sense, one could call the primary branch the ``physical'' one, although we will not use such a term\footnote{
The reason is that this would imply that the secondary branch is ``unphysical'', which we are reluctant to claim. Although the secondary branch definitely does not have direct interpretation in terms of ``physics'' of the input S-matrix, it might very well have some physical content of its own. In fact, understanding physical mechanism behind the secondary branch is one of the outstanding problems which remains open both for the generalized TTbar deformations and for the TTbar proper.
}.
The secondary branch always has lower energy $E(R)$ than the primary one, which is qualitatively similar to the behavior observed in the TTbar deformations with negative $\alpha$, see Fig \ref{ERplotTTbar}. Since the two branches merge at some finite $R=R_*$, this can be regarded as a ``turning point'', where the continuation along the graph of $E(R)$ turns backward into the secondary branch. This is precisely the kind of situation the PALC method is designed to deal with. The secondary branch remains real for all $R>R_*$ and, moreover, develops a linear asymptotic $\sim e_{\infty}\, R$ as $R\to \infty$. This, again, is in full qualitative agreement with the TTbar deformations, together with the important fact that the singularity of the pseudo-energy $\epsilon(\theta|R)$, viewed as a function of $R$, occurs at $R=R_*$ that is independent of $\theta$. Of the above features, the existence of primary and secondary branches with the turning point at finite $R_*$, independent from $\theta$, repeat \emph{verbatim} in the bosonic 2CDD model. On the other hand, we still cannot check the large $R$ behavior of the secondary branch with sufficient accuracy, due to some instability in the numerical procedure. We will return on this problem in a future work.

It is likely that the general situation displayed in the models studied here, i.e. the solution of the TBA equation developing a square-root singularity at finite $R_*$, which signals the presence of a Hagedorn transition, remains qualitatively the same when more CDD poles are added in \eqref{phipole} -- with the possible exceptions of special domains, hypersurfaces of lower dimension, in the parameter space\footnote{Examples of such cases can be found in \cite{Martins:1992ht,Martins:1992yk,Dorey:2000zb}.}. This of course will have to be carefully verified. We regard the present work as a first step in the program of systematically studying the short-distance behavior of the generalized TTbar deformations \eqref{Aalphas} of IQFTs. The qualitative similarity to the TTbar-deformed QFTs, with negative $\alpha$, suggests that the same mechanism behind the formation of the Hagedorn singularities is at play in all of these models. Understanding the physics underlying this phenomenon remains the most important open problem in this context, as well as the main motivation for the present work.

\section{From TBA to Hagedorn: the TTbar case}\label{sec:TTbar}

Henceforth we will assume that the theory under consideration is integrable, with a factorizable $S$-matrix. Let us briefly remind how, in this case, equation \eqref{burgers2} can be derived from the $S$-matrix deformation \eqref{sdef} via the TBA equations. We will present a somewhat simplified version of the much more general arguments of \cite{Cavaglia:2016oda} 
(for related work see \cite{Dubovsky:2012wk,Caselle:2013dra} and the more recent \cite{LeClair:2021wfd,LeClair:2021opx}).
Whereas the analysis in \cite{Cavaglia:2016oda} applies to all the energy eigenvalues of the TTbar deformed theory \eqref{Aalpha}, we limit our considerations to the ground-state energy, which we denote as $E(R)$. The advantage is that the simple arguments presented below apply to the deformation \eqref{sdef} of an essentially generic integrable theory. The only assumptions, made for simplicity, are that the particle scattering theory associated with $\mathcal{A}_0$ involves only one kind of neutral particles, with the factorizable scattering of fermionic type\footnote{Extension to the bosonic case $S(0)=+1$ is trivial. Less straightforward but still possible is the generalization to the cases of a scattering theory involving many species of particles,including the bound states, with different or equal masses. We will elaborate on such cases elsewhere.}, i.e. $S_0(0)=-1$. The goal is to emphasize some important properties of the solution which, as we will see, are shared by the TBA solutions by more general CDD deformations.

The TBA equation \eqref{tbas} associated with the deformed $S$-matrix \eqref{sdef} has the following kernel
\begin{eqnarray}\label{varphialpha}
\varphi_\alpha (\theta-\theta') = \varphi_0 (\theta-\theta') - \alpha\,\cosh(\theta-\theta')\;.
\end{eqnarray}
Recall that the ground state energy $E_\alpha(R)$ is given by \eqref{etbas}, which in our case reads
\begin{eqnarray}
E_\alpha (R)= - \,\int_{-\infty}^{\infty} \,\cosh\theta\,\,L_\alpha(\theta|R)\,\frac{d\theta}{2\pi}\;,
\label{eq:enalpha}
\end{eqnarray}
where $L_\alpha (\theta|R):=\log\left(1+e^{-\epsilon_\alpha (\theta|R)}\right)$ satisfies the deformed TBA equation \eqref{tbas},
\begin{eqnarray}
\epsilon_\alpha(\theta|R) = R\, \cosh\theta - \int\, \varphi_\alpha(\theta-\theta')\,L_\alpha(\theta'|R)\,\frac{d\theta'}{2\pi}\;.
\end{eqnarray}
Due to the fact that the pseudo-energy is even, as is easily shown, we can separate the dependence on $\theta$ and $\theta'$ in the rightmost term in the kernel \eqref{varphialpha} so that the TBA equation can be written as follows
\begin{eqnarray}
\epsilon_\alpha(\theta|R) = \left(R-\alpha\,E_\alpha(R)\right)\,\cosh\theta - \int\,\varphi_0(\theta-\theta')\,L_\alpha(\theta'|R)\,\frac{d\theta'}{2\pi}\;,
\label{ttbartba2}
\end{eqnarray}
where we used the definition \eqref{eq:enalpha}.
For reasons that will become clear shortly we have made explicit the fact that $\epsilon(\theta|R)$ and $L(\theta|R)$ are functions of $R$ as well as of the rapidity $\theta$. This last form \eqref{ttbartba2} shows that $\epsilon_\alpha(\theta|R)$ satisfies the same TBA equation as $\epsilon_0(\theta|R)$, only with $R$ replaced by
$R-\alpha E_\alpha(R)$. It then follows that
\begin{eqnarray}
\epsilon_\alpha(\theta|R) = \epsilon_0(\theta|R-\alpha E_\alpha(R))
\end{eqnarray}
which immediately implies the equation \eqref{burgers2} for the deformed energy.

\begin{figure}[ht]
\centering
    \includegraphics{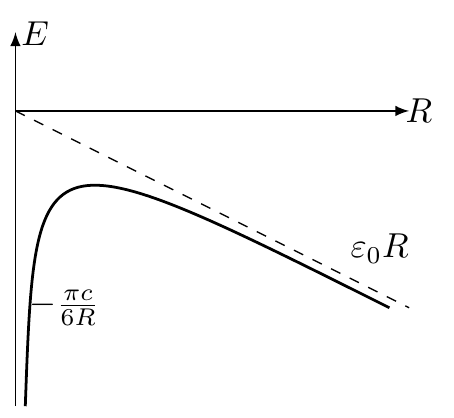}
\caption{\small{ {Finite-size ground state energy $E_0(R)$ of a conventional Wilsonian relativistic QFT. Its $R\to 0$ behavior $-\pi c/6R$ is controlled by the UV fixed point. At large $R$, $E_0(R)$ shows the linear behavior $\simeq \varepsilon_0 R$, with the slope $\varepsilon_0$ representing the bulk vacuum energy density. We have to stress that the TBA equations actually compute the difference $E_\text{vac}(R)-\varepsilon_{0} R$, and in our subsequent analysis $E(R)$ stands for this difference. (That is why in all plots below the $R\to\infty$ slope of the primary branch is always set to zero.)}}}\label{EvacQFT}
\end{figure}

It is also worth reminding here how the singularity of $E_\alpha(R)$, signifying the Hagedorn density of states, follows from \eqref{burgers2}. This takes a particularly simple form in terms of the function $R_\alpha(E)$, inverse to the function $E_\alpha(R)$, where $\alpha$ is regarded as a fixed parameter,
\begin{eqnarray}\label{burgers3}
R_\alpha(E)=R_0(E)+\alpha\,E\;.
\end{eqnarray}
This expression shows that the graph of the deformed function $E_\alpha(R)$ differs from the graph of $E_0(R)$ just by an affine transformation $(R,E) \to (R+\alpha E, E)$ of the $(R,E)$ plane. If we assume, as we do, that the undeformed theory $\mathcal{A}_0$ is a conventional QFT, defined \`{a} la Wilson as the RG flow from some UV fixed point down to an IR one (see \cite{Wilson:1973jj}), then the graph of $E_0(R)$ looks qualitatively as shown in Fig \ref{EvacQFT}.

\begin{figure}[ht]
\centering
    \begin{subfigure}{4cm}
      \includegraphics{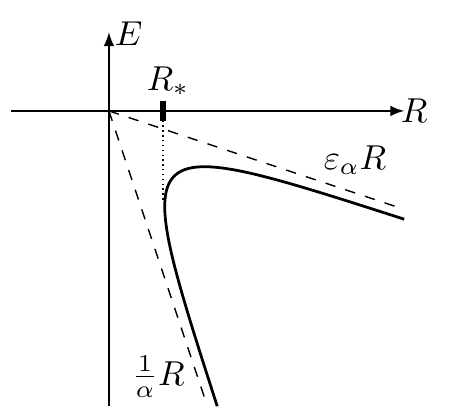}
      \caption{$\alpha<0$}
    \end{subfigure}
    \hspace{2cm}
    \begin{subfigure}{4cm}
      \includegraphics{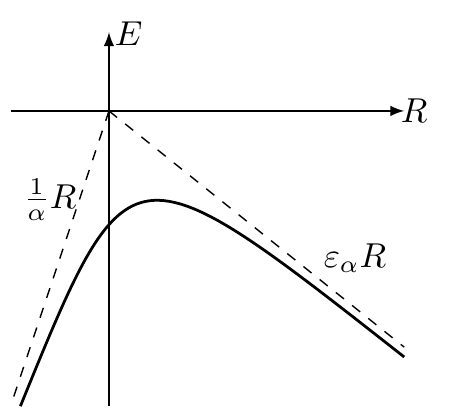}
      \caption{$\alpha>0$}
    \end{subfigure}
\caption{\small{Finite-size ground state energy of the TTbar deformed theory. (a) $\alpha <0$. The graph $E_\alpha(R)$ shows the ``turning point" at some finite $R_*$, which signals the Hagedorn transition. (b) $\alpha >0$. $E_\alpha(R)$ shows no singularity at $R=0$.}}\label{ERplotTTbar}
\end{figure}

At large $R$ the function $E_0(R)$ approaches a linear asymptotic
$\varepsilon_0 R$, where $\varepsilon_0$ is the vacuum energy density of the infinite system, with the rate of the approach controlled by the IR fixed point, which, typically, is a non-critical one. On the other hand, at $R\to 0$ it diverges as the Casimir energy determined by the UV fixed point, $E_0(R) \to -\pi c/6R$, where $c$ is the Virasoro central charge of the UV fixed point CFT. Then, according to \eqref{burgers3}, the plot of $E_\alpha(R)$ will look as either one of the panels \emph{a}) or \emph{b}) in Fig \ref{ERplotTTbar}, depending on the sign of $\alpha$. In what follows we will concentrate our attention to the case of negative $\alpha$, shown in panel \emph{a}). Note that the curve $E_\alpha(R)$ has two branches, each of which having real values for $R$ above a certain critical value $R_*$. 
It is the upper ``primary'' branch that corresponds to the ground state energy of the TTbar-deformed theory
\eqref{Aalpha}.

The two branches merge at $R=R_*$, where the function $E_\alpha (R)$ develops a square-root branch point, i.e. the derivative $dE_\alpha(R)/dR$ diverges as $(R-R_*)^{-1/2}$. At $R<R_*$, the analytic continuation of $E_\alpha(R)$ returns complex values and the two branches are complex conjugate.

\begin{figure}[ht]
\centering
    \includegraphics{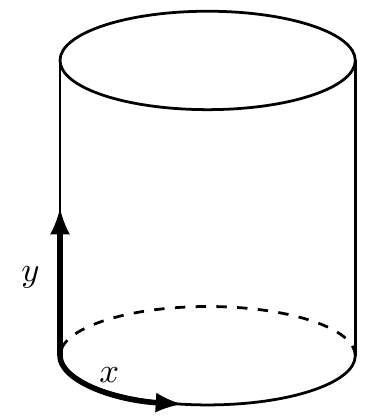}
\caption{\small{{The Euclidean space-time cylinder representing the finite-size geometry 
in our analysis. The coordinate $x$ is compactified on a circle of circumference $R$, while the length $L$ of the cylinder is assumed to be asymptotically large. In the
picture where $y$ is regarded as the Euclidean ``time" the partition function \eqref{eq:logZ} is dominated by the finite-size ground state contribution. In the complementary picture, where $y$ is interpreted as spatial coordinate while $x$ plays the role of the Matsubara ``time", the same partition function is given by the thermal trace \eqref{eq:Ztrace}.}}}\label{cylinder}
\end{figure}

It is the singularity at $R_*$ that signals the Hagedorn phenomenon in the deformed theory, which can be inferred as follows. When the Euclidean theory is considered in the geometry of a very long cylinder of circumference $R$, as shown in Fig \ref{cylinder}, its partition function $Z$ is saturated by the finite-size ground state 
\begin{eqnarray}\label{eq:logZ}
-\log Z \ \simeq L\,E_\alpha(R)\;,
\end{eqnarray}
where $L\to \infty$ is the length of the cylinder. This corresponds to the picture in which the coordinate $y$ along the cylinder is taken as the Euclidean time. Alternatively, if one uses the picture where $x$ plays the role of Matsubara time, the same partition function is represented as the trace
\begin{eqnarray}\label{eq:Ztrace}
Z = \text{tr}\left(e^{-R {\hat H}_x}\right) = \int_0^\infty \,d{\cal E}\,\,\Gamma({\cal E})\,e^{-R{\cal E}} = e^{-RF(R)}
\end{eqnarray}
where $d{\cal E} \Gamma({\cal E})\sim d{\cal E}\,e^{{\cal S}({\cal E})}$ denotes the density of states, i.e. the number of states in the energy interval $d{\cal E}$. While in a local QFT whose high-energy limit is governed by the UV fixed point the entropy ${\cal S}$ grows as ${\cal S}({\cal E}) \simeq \sqrt{2\pi c/3}\ \sqrt{L {\cal E}}$ as ${\cal E}\to \infty$ -- this is known as Cardy formula \cite{Cardy:1986ie} -- the singularity of $F(R)$ at finite positive $R=R_*$ is formed when the entropy ${\cal S}({\cal E})$ grows much faster,
\begin{eqnarray}\label{hagedorn1}
{\cal S}(\cal E) \simeq R_* \,{\cal E}\,,
\end{eqnarray}
so that the partition sum diverges at $R<R_*$. We will discuss the density of states in the TTbar deformed theories in more details in \S \ref{sec:discussion}.

In the above discussion we have denoted by $R_*$ the position of the singularity of $E_\alpha(R)$ as a function of $R$. It is important to observe that the solution $\epsilon_\alpha (\theta|R)$ displays a singularity at the same position $R=R_*$, independent on the value of the rapidity $\theta$. In other words, in the two-dimensional space spanned by the variables $(\theta, R)$ the singularity of $\epsilon_\alpha (\theta|R)$ occurs along the line $(\theta, R=R_*)$. We will see that this feature of the singularity associated with the Hagedorn transition will be reproduced in the generalized TTbar flows studied below.

As already mentioned, the enhancement of the density of states in the deformed theory is well expected. The scattering phase in \eqref{sdef} grows fast with the center of mass energy, leading to the increase of the density of two-particle states, implying a yet greater increase of the density of all multi-particle states. The calculation presented above demonstrates that the resulting entropy displays the Hagedorn behavior \eqref{hagedorn1}. It is then tempting to assume that the formation of the Hagedorn density \eqref{hagedorn1} is directly related to the fast growth of the scattering phase. In the next section we will show that the Hagedorn singularity develops just as well in models whose associated CDD factor has a finite behavior at high energies, as in \eqref{phipole} with finite $N$, which indicates that the physical origin of the Hagedorn transition in the deformed theories is substantially more intricate. 

\section{The models}\label{sec:models}

Here we study the CDD deformations of the trivial (fermionic or bosonic) $S$-matrix by the pole factor \eqref{phipole}, which we write as
\begin{eqnarray}\label{sNpole}
S(\theta) = \sigma\,\prod_{p=1}^N \,\frac{i\sin u_p+\sinh\theta}{i\sin u_p-\sinh\theta}
\end{eqnarray}
where, as before, $\sigma=-$ (resp. $\sigma=+$) corresponds to the fermionic (resp. bosonic) case. The parameters $u_p$ may be taken to be complex and, in view of the obvious periodicity of $S(\theta)$, we may limit our attention to the strip $-\pi < \Re(u_p) < \pi$. The standard analytic requirements for the physical $S$-matrix, however, impose restrictions on the possible locations of the poles $\theta_p = iu_p$. Taking these restrictions into consideration, the parameters $u_p$ are allowed to be either real or complex with negative real parts. The poles $\theta_p = i u_p$, with real positive $u_p$ signal the existence of bound states -- new stable particles of mass $2M\,\cos(u_p/2)$. Since the presence of such particles violates our working assumption that the mass spectrum of the theory only involves a single kind of stable particle with mass $M$, henceforth we will assume that all parameters $u_p$ in \eqref{sNpole} possess a negative real part\footnote{This leaves out the possibility of having a pole at $\theta=2\pi i/3$ which may be identified with the same particle of the mass $M=1$. Such interpretation requires that $S(\theta)$ satisfies additional bootstrap condition. This possibility, known as the ``$\varphi^3$ property", cannot be realized in the 2CDD model considered in this work, but may be relevant when $N$ is greater than 2. We hope to address this type of models elsewhere.
}:
\begin{eqnarray}
	-\pi \leq \Re(u_p)\leq 0\;,\qquad \forall p=1,\ldots ,N\;.
\label{eq:up_analiticity}
\end{eqnarray}
This leaves us with poles $\theta_p = i u_p$ lying in the unphysical region, i.e. the region of the complex center-of-mass energy $s$-plane reached by analytically continuing the scattering amplitude through the two-particle branch cut.
When $u_p$ has nonzero imaginary parts, such poles are associated to unstable particles, having complex masses $M_p = 2M\,\cos(u_p/2)$, with the real and imaginary parts identified as usual with the mean center of mass energy and the width of the resonances. The poles with real negative $u_p$ do not have clear particle interpretation, but the number of such poles signify the increment of the scattering phase as the function of $\theta$ at low energies; these poles are often referred to as the virtual states (see e.g.\cite{perelomov1998quantum}).

A final requirement is that of unitarity of the physical $S$-matrix, which demands that $S(-\theta) = S^* (\theta)$ at all real $\theta$, or, equivalently, that $S(\theta)$ takes real values at pure imaginary $\theta$. It follows that any non-real parameter $u_p$ in \eqref{sNpole} either has fixed real part $\Re(u_p)=-\pi/2$ or appears together with its conjugate $u_p^{\ast}$. We can then refine the range (\ref{eq:up_analiticity}) to the following three cases
\begin{eqnarray}
	\textrm{a})&\quad&\Im(u_p) = 0\quad\textrm{and}\quad \Re(u_p)\in\left(-\pi,0\right)\;, \nonumber\\
	\textrm{b})&\quad&\Im(u_p) \neq 0 \quad\textrm{and}\quad\Re(u_p)=-\frac{\pi}{2}\;,
\label{eq:up_analiticity_refined}\\
	\textrm{c})&\quad&\Im(u_p) >0 \quad\textrm{and}\quad \Re(u_p)\in\left(-\pi,-\frac{\pi}{2}\right)\cup\left(-\frac{\pi}{2},0\right] \nonumber\\
	\phantom{\textrm{c})}&\quad&\phantom{\Im(u_p) >0}\quad \textrm{and}\quad \exists p'\leq N\quad \textrm{s.t.}\quad u_{p'} = u_p^{\ast}\;. \nonumber
\end{eqnarray}
Thus, each subfamily $(\sigma,N)$ of \eqref{sNpole} contains a number of, in principle, different models, determined by a given combination of the ranges \eqref{eq:up_analiticity_refined} for each of the parameters $\lbrace u_p\rbrace_{p=1}^N$. Some simple combinatorics\footnote{
Given the number $N$ of poles one needs to partition it into three non-negative integers $n_a$, $n_b$ and $n_c$ with the constraint that $n_a+n_b+2n_c=N$. Once a value of $n_c = 0,1,\ldots, \lfloor N/2\rfloor$ is chosen, one is obviously left with $N-2n_c+1$ non-equivalent arrangements of poles between the cases a) and b). Thus the number of different models is given by $\sum_{n_c=0}^{\lfloor N/2\rfloor}(N-2n_c+1)$, which gives the result \eqref{eq:number_of_models}.
} tells us that this number is 
\begin{eqnarray}
	\frac{1}{4}N^2 + N +\frac{7+(-1)^N}{8} = \left\lbrace \begin{array}{l l l} \left(\frac{N}{2}+1\right)^2 & & N\in2\mathbb Z_{>0} \\ \\ \frac{N+1}{2}\frac{N+3}{2} & & N\in2\mathbb Z_{>0}-1 \end{array} \right. \;.
\label{eq:number_of_models}
\end{eqnarray}

Since for any model determined by \eqref{sNpole}, with parameters in the ranges \eqref{eq:up_analiticity}, the mass spectrum contains a single stable excitation, the resulting single-particle TBA equation takes the simple form (\ref{tbas}), with the kernel $\varphi(\theta)$ being the derivative of the scattering phase which, in the case of \eqref{sNpole}, explicitly reads
\begin{eqnarray}
	\varphi_{N\textrm{CDD}}(\theta) = \frac{1}{i}\frac{\partial}{\partial \theta} \log S_{N\textrm{CDD}}(\theta) = - \sum_{p=1}^N \frac{2\sin u_p\,\cosh\theta}{\sin^2 u_p +\sinh^2\theta} \;.
\label{eq:TBA_kernel}
\end{eqnarray}
An equivalent, sometimes more useful, expression of this kernel is its partial fractions expansion 
\begin{eqnarray} \label{eq:kernelNCDD}
\varphi_{N\textrm{CDD}}(\theta) = \sum_{p=1}^N\left[\frac{1}{\cosh\left(\theta+i(u_p+\frac{\pi}{2})\right)} +  \frac{1}{\cosh\left(\theta-i(u_p+\frac{\pi}{2})\right)}\right]\;.
\label{eq:TBA_kernel_v2}
\end{eqnarray}
In what follows, we are going to concentrate our attention on two particular subfamilies: the ``1CDD models'' where $N=1$ and the ``2CDD models'' with $N=2$.

\paragraph{The 1CDD models}

When $N=1$ the $S$-matrix \eqref{sNpole} consists of a single factor
\begin{eqnarray}
S_{1\textrm{CDD}}(\theta) = \sigma\,\frac{i\sin u_1+\sinh\theta}{i\sin u_1-\sinh\theta}\,.
\label{1cdd}
\end{eqnarray}
According to the breakdown of cases \eqref{eq:up_analiticity_refined}, for each choice of the TBA statistics we only have two possible models, corresponding to the following ranges of the parameter $u_1$:
\begin{enumerate}[label=(\alph*)]
	\item $u_1\in\mathbb{R}$ and $-\pi<u_1<0$,
	\item $u_1 = -\pi/2 + i \theta_0$ and $\theta_0 \in\mathbb R$.
\end{enumerate}
Considering at first the fermionic case $\sigma=-1$, one recognizes in \eqref{1cdd}, for the case (a), the well-known $S$-matrix of the sinh-Gordon model
\begin{eqnarray}
S_{\textrm{shG}}(\theta) = -\frac{i\sin u_1+\sinh\theta}{i\sin u_1-\sinh\theta}\,,\qquad -\pi<u_1<0\;.
\label{eq:shG_Smat}
\end{eqnarray}
On the other hand, the case (b) corresponds to the $S$-matrix of the ``staircase model'', introduced in \cite{Zamolodchikov:1991pc}
\begin{eqnarray}
S_{\textrm{stair}}(\theta) = \frac{\sinh\theta-i\cosh\theta_0}{\sinh\theta+i\cosh\theta_0}\;,\qquad \theta_0 \in\mathbb R\;.
\label{eq:stair_Smat}
\end{eqnarray}
In both the cases (a) and (b) of the fermionic 1CDD model, the iterative solution to the TBA equation converges at all positive values of $R$, producing a function $E(R)$ analytic in the half-line $R>0$ and displaying a Casimir-like singularity at $R=0$, in full agreement with the interpretation of $E(R)$ as the ground state energy of a UV complete local QFT.

For what concerns the two bosonic 1CDD models, the solution of the TBA equation has a considerably more intricate behavior. The case (a) of $u_1$ real was first addressed in \cite{Mussardo:1999aj}, where it was observed that the iterative solution of the TBA equation only converges for sufficiently large radius $R>R_*>0$. The authors also noticed that the function $E(R)$ appears to develop some sort of singularity at $R=R_*$. Below in \S \ref{sec:results} we will show that the solution to the TBA equation, and, consequently, the ground state energy $E(R)$, possesses, as a function of $R$, two branches. These merge at $R=R_*$, meaning that $R_*$ is a square-root branching point. We also show that this behavior extends to the case (b) of complex parameter $u_1=-\pi/2+i\theta_0$.

\paragraph{The 2CDD model}
In the $N=2$ subfamily, a pair of CDD factors is present in \eqref{sNpole}:
\begin{eqnarray}
	S_{\textrm{2CDD}}(\theta) = \sigma\,\frac{i\sin u_1 + \sinh\theta}{i\sin u_1 - \sinh\theta}\frac{i\sin u_2 + \sinh\theta}{i\sin u_2 - \sinh\theta}\;.
\label{eq:2cdd_general}
\end{eqnarray}
Following the breakdown \eqref{eq:up_analiticity_refined}, we see that there are $4$ possibly distinct models, corresponding to the following ranges of the parameters $u_1$ and $u_2$
\begin{enumerate}[label=(\alph*)]
	\item $u_1\in\mathbb{R}$ and $-\pi<u_1<0$,\\ $u_2\in\mathbb{R}$ and $-\pi<u_2<0$,
	\item $\theta_0 \in\mathbb R$ and $u_1 = -\pi/2 + i \theta_0$, \\$u_2 \in\mathbb R$ and $-\pi<u_2<0$,
	\item[(b')] $u_1 \in\mathbb R$ and $-\pi<u_1<0$, \\$\theta_0 \in\mathbb R$ and $u_2 = -\pi/2 + i \theta_0$,
	\item $\theta_0 \in\mathbb R$ and $u_1 = -\pi/2 + i \theta_0$, \\$\theta_0' \in\mathbb R$ and $u_2 = -\pi/2 + i \theta_0'$,
	\item $\theta_0 \in \mathbb R$, $\gamma \in (-\pi/2,\pi/2)$, $u_1 = \gamma - \pi/2 + i \theta_0$ and $u_2 = u_1^{\ast}$.
\end{enumerate}
The model (a) can be considered as a special instance of the more general case d). On the other hand the models (c) and (b) -- equivalent to (b') -- are genuinely distinct. All the models above display, both for the bosonic and fermionic statistics, the same type of behavior observed in the bosonic 1CDD models mentioned above: the iterative procedure for solving the TBA equation \eqref{tbas} only converges for $R$ larger than some positive value $R_\ast>0$ and the ground state energy $E(R)$ apparently develops a singularity at $R=R_\ast$.

While we are going to present some data for all the various 2CDD cases, we devoted most of our attention to the case (d), that we will call, with some definitional abuse, the ``2CDD model''. Its $S$-matrix and TBA kernels explicitly read as follows
\begin{eqnarray}
S_{\textrm{2CDD}}(\theta) = \sigma\, \frac{\sinh\theta - i\cosh(\theta_0+i\pi\gamma)}{\sinh\theta + i\cosh(\theta_0+i\pi\gamma)}\frac{\sinh\theta - i\cosh(\theta_0-i\pi\gamma)}{\sinh\theta + i\cosh(\theta_0-i\pi\gamma)}\;.
\label{eq:2cdd_particular}
\end{eqnarray}

\begin{eqnarray}
	\varphi_{\textrm{2CDD}}(\theta) = \sum_{\eta,\eta'=\pm}\frac{1}{\cosh(\theta+\eta\theta_0+i\eta'\gamma)} \;.
\label{eq:2cdd_kernel}
\end{eqnarray}

\subsection{Iterative solution}\label{subsec:iterative}

The chances of a non-linear integral equation of the form \eqref{tbas} to be amenable to an explicit analytic solution are considerably slim. For this reason the main investigation approach to the TBA equations is of numerical nature\footnote{In some limiting cases, it is possible to derive exact expressions, e.g. for the ground-state energy in the conformal limit, via the so-called ``dilogarithm trick'', as explained nicely in \cite{Fendley:1993jh}.}. In most situations, a simple iterative procedure of the following type
\begin{eqnarray}
	\epsilon_n(\theta) = R \cosh\theta + \sigma \intop \varphi(\theta - \theta') \log\left[1-\sigma e^{-\epsilon_{n-1}(\theta')}\right]\,\frac{d\theta}{2\pi}\;,
\label{eq:iterative_routine}
\end{eqnarray}
appropriately discretized, is shown to converge to the actual solution
\begin{eqnarray}
	\lim_{n\rightarrow\infty} \epsilon_n(\theta) = \epsilon(\theta)\;,
\label{eq:limit_solution_iteration}
\end{eqnarray}
when the seed function $\epsilon_0(\theta)$ is chosen as the driving term\footnote{In the case in which the iterative procedure does converge, there is actually a vast freedom in the choice of the seed function. However the standard choice indicated in the main text is the most natural one.}
\begin{eqnarray}
	\epsilon_0(\theta) = R\cosh\theta\;.
\end{eqnarray}
The existence and uniqueness of the limit \eqref{eq:limit_solution_iteration} has been proven rigorously in \cite{Fring:1999mn} for the fermionic single-particle\footnote{See also \cite{Hilfiker:2017jqg} for an extension to fermionic multi-particle TBA equations.} TBA equation \eqref{tbas} with a kernel satisfying the requirement
\begin{eqnarray}
	\left\vert\left\vert\varphi\right\vert\right\vert_1 := \intop\, \left\vert\varphi(\theta)\right\vert \frac{d\theta}{2\pi} \leq 1\;.
\end{eqnarray}
The fermionic 1CDD models do satisfy this condition and, as such, the iteration procedure is guaranteed to converge nicely in the whole range $R\in\mathbb R_{>0}$, a fact which is easily verified numerically. All the other models we considered above, on the other hand, violate one or more of the hypotheses of the existence and uniqueness theorem in \cite{Fring:1999mn} -- being either of bosonic statistic, or having a kernel with $L^1$ measure $\vert\vert\varphi\vert\vert_1=2$, or both -- and are not guaranteed to possess a convergent iterative solution. Notice that the $L^1$ measure of the TBA kernel \eqref{eq:TBA_kernel_v2} counts the number of CDD factors
\begin{eqnarray}
	\left\vert\left\vert \varphi_{N\textrm{CDD}} \right\vert\right\vert_1 = N\;,
\label{eq:L1_kernel}
\end{eqnarray}
meaning that, in the class of models described by the $S$-matrix \eqref{sNpole}, only the subfamily with $(\sigma,N) = (-1,1)$ is guaranteed to have a convergent iterative solution.

\begin{figure}[h!]
\begin{center}
\includegraphics{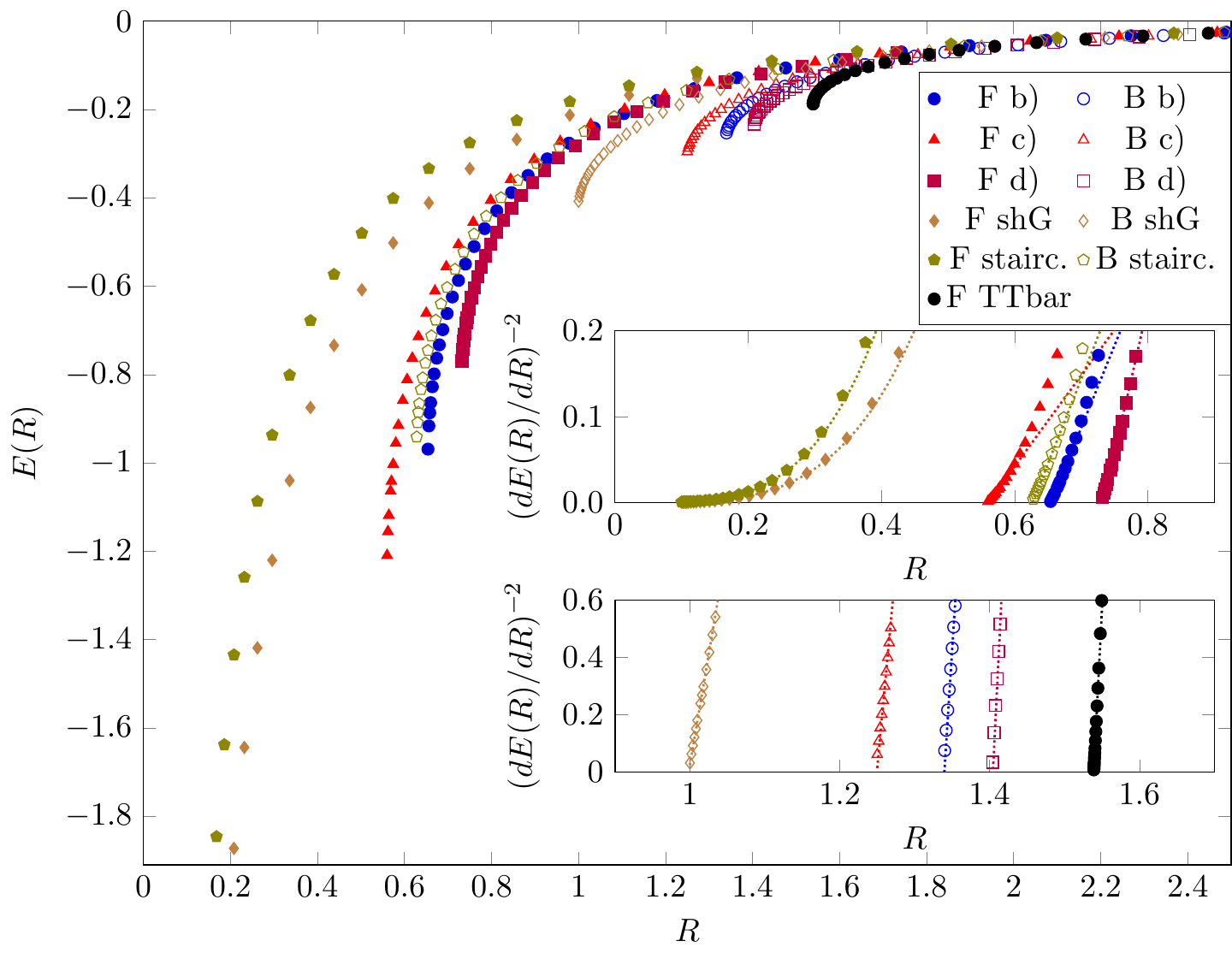}
\end{center}
\caption{Ground-state energies for the various models discussed above, along with that of the $T\bar{T}$-deformed free fermion (black dots). The empty (resp. filled) markers correspond to models with bosonic (resp. fermionic) statistics. The fermionic sinh-Gordon and staircase models can be solved iteratively all the way to the $R\to 0$ limit, while the rest fail to converge below a certain model-specific scale $R_*$. The parameters of the models were chosen as to allow a comfortable visual comparison between the curves and are the same for both bosonic and fermionic versions of the same model. Insets: inverse square of the (numerical) derivative. As shown by the fits (dotted lines), the fermionic sinh-Gordon and staircase models show the conventional UV behavior $\propto R^4$, while the other models develop a $\propto R$ behavior reminiscent of the square-root branching singularity of the ground state energy. }
\label{fig:2CCDmodelscomp}
\end{figure}

We investigated numerically the 1CDD models (a) and (b) and the 2CDD models (b) to (d)\footnote{Remember that the 2CDD model (a) is really a sub-case of model (d).}, for both the bosonic and fermionic statistic, using the iterative procedure \eqref{eq:iterative_routine}. As already mentioned above we observed that only for the 1CDD fermionic models this procedure converges for all positive values of the radius $R$. In every other case, there exists a positive ``critical radius'' $R_{\ast}>0$ such that for $R\leq R_{\ast}$ the iterative routine stops converging. As $R$ approaches $R_{\ast}$ from larger values, we noticed that the rate of convergence of the iterative numerical routine slows down dramatically, a telltale sign of the existence of some kind of singularity nearby\footnote{This same ``critical slowing down'' of the numerical iterative procedure is observed as $R\rightarrow 0$ in any TBA system with iterative solution converging in $R\in\mathbb R_{>0}$. In this cases it reflects the existence of a Casimir-like singularity of the ground-state energy at $R=0$.}. In Figure \ref{fig:2CCDmodelscomp} we collected the plots of the ground-state energy $E(R)$ for one representative point in the parameter space for each of the models we mentioned above along with one for the $T\bar{T}$-deformed free fermion. The shape of the curves suggests that all the cases, apart from the fermionic 1CDD models, behave qualitatively in the same way as the $T\bar{T}$-deformed free fermion, that is to say they develop a square-root type singularity at some critical value of the radius $R=R_{\ast}>0$:
\begin{eqnarray}
	E(R) \underset{R\rightarrow R_{\ast}^+}{\sim} c_0 + c_{1/2}\sqrt{R-R_{\ast}}+\mathcal{O}(R-R_{\ast})\;.
\label{eq:supposed_square_root_behaviour}
\end{eqnarray}
In order to further confirm this suspicion we plotted the derivative of the ground-state energy to the power $-2$ in the vicinity of the supposed critical point. As we can see in the insets of Figure~\ref{fig:2CCDmodelscomp}, the numerical results are in good accord with the hypothesis that $R_{\ast}$ is a singular point of square root type, as expressed by \eqref{eq:supposed_square_root_behaviour}.

\subsection{Two branches}\label{subsec:two_branches}

Having our expectation confirmed leaves us with the question of how to deal numerically with such a square root critical point. In particular, the behavior \eqref{eq:supposed_square_root_behaviour} implies the existence of a secondary branch of the ground-state energy, behaving as
\begin{eqnarray}
	\tilde{E}(R) \underset{R\rightarrow R_{\ast}^+}{\sim} c_0 - c_{1/2}\sqrt{R-R_{\ast}}+\mathcal{O}(R-R_{\ast})\;,
\label{eq:supposed_square_root_behaviour_second_branch}
\end{eqnarray}
in the vicinity of the critical point. Here and below we are going to use the notation $\tilde{E}(R)$ for the secondary branch. We would like to be able to access numerically to this secondary branch and to explore its properties, e.g. its large $R$ behavior and the possible existence of further critical points. The iterative routine \eqref{eq:iterative_routine} is ill suited for this job and we need to employ a more refined method, the PALC mentioned in the introduction and described in \S \ref{sec:num_meth}. Deferring a more thorough analysis of the properties of $E(R)$ to \S \ref{sec:results}, let us present here its main qualitative features, concentrating on a single point in the parameter space of the fermionic 2CDD model (d) as a representative case.

\begin{figure}[t!]
\begin{center}
\includegraphics{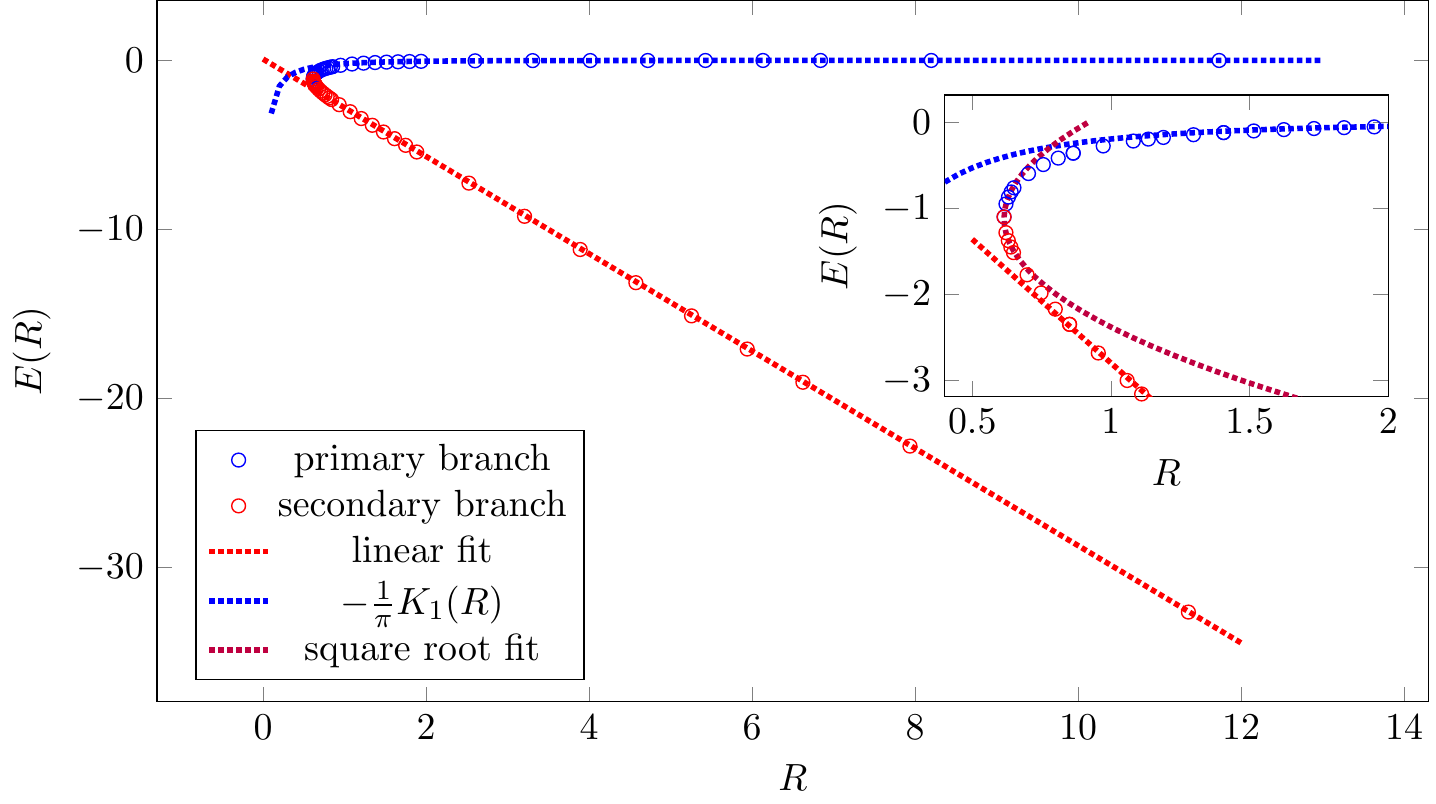}
\end{center}
\caption{Here is plotted the ground-state energy $E(R)$ for the model with $S$-matrix \eqref{eq:2cdd_particular} with $\theta_0 = 1/2$ and $\gamma = 3\pi/20$, obtained through the PALC routine described in \S \ref{sec:num_meth}. The numerical points are sided by three lines, approximating $E(R)$ for large $R$ on both branches and for $R\gtrsim R_{\ast}$.}
\label{fig:both_branches}
\end{figure}
More specifically let us set $\theta_0 = 1/2$ and $\gamma = 3\pi/20$ and compute numerically the ground-state energy of the model defined by the $S$-matrix \eqref{eq:2cdd_particular}. The result is displayed in Figure \ref{fig:both_branches}. We see that the function $E(R)$ does indeed possess two branches with distinctly different IR behavior. The primary branch is characterized by the universal IR behavior
\begin{eqnarray}
	E(R)\underset{R\rightarrow\infty}{\sim} - \frac{1}{\pi}\,K_1(R) + \mathcal{O}\left(e^{-2R}\right)\;,
\label{eq:E_asymp_primary}
\end{eqnarray}
where $K_1$ stands for the modified Bessel function while the secondary branch approaches a linear behavior at large $R$
\begin{eqnarray}
	\tilde{E}(R)\underset{R\rightarrow\infty}{\sim} - \varepsilon_{-} R \;,
\label{eq:E_asymp_secondary}
\end{eqnarray}
with a rate of approach likely to be some negative power of $R$.
For the specific case depicted in Figure \ref{fig:both_branches} the coefficient of the linear term is found to be
\begin{eqnarray}
	\varepsilon_{-}\left(\theta_0 = \frac{1}{2},\gamma = \frac{3\pi}{20}\right) = -2.87452\ldots\;,
\end{eqnarray}
while the constant term is vanishing up to the precision we used for our numerical routines. We will see in \S \ref{sec:results} that this is the asymptotic behavior predicted by analytical considerations. In the zoomed box in Figure \ref{fig:both_branches} we also plotted a fit of the function $E(R)$ in the vicinity of the critical point $R_{\ast}$. As expected the behavior in this region is best described by the square-root function \eqref{eq:supposed_square_root_behaviour} (and \eqref{eq:supposed_square_root_behaviour_second_branch} for the secondary branch), with the coefficients taking the following values
\begin{eqnarray}
	c_0\left(\theta_0 = \frac{1}{2},\gamma = \frac{3\pi}{20}\right) &=& -1.11767\ldots\;, \nonumber \\
	c_{1/2}\left(\theta_0 = \frac{1}{2},\gamma = \frac{3\pi}{20}\right) &=& 2.03547\ldots\;, \\
	R_{\ast}\left(\theta_0 = \frac{1}{2},\gamma = \frac{3\pi}{20}\right) &=& 0.61478849\ldots\;.\nonumber
\end{eqnarray}
Another notable fact is that we see no trace of additional singular points: the PALC method can, apparently, reach arbitrarily large values of $R$ on the secondary branch and the resulting ground-state energy quickly approaches the expected asymptotic linear behavior.

We note again that the behavior of $E(R)$ depicted in Figure \ref{fig:both_branches} is qualitatively identical to the one exhibited by the ground-state energy of $T\bar{T}$-deformed models for negative values of the deformation parameter $\alpha$, as described in \S \ref{sec:TTbar} (see e.g. Figure \ref{ERplotTTbar}).

Finally, we stress that the features of $E(R)$ described here for a point in the parameter space of a specific model really are representative of the general behavior of the ground-state energy in the family of models defined by the S-matrices \eqref{sNpole}, at least for what concerns the case of fermionic statistics. As we will discuss in \S \ref{sec:results} the status of the models with bosonic statistics is still not completely settled. In particular it is still unclear whether the secondary branch of $E(R)$ displays additional critical points or continues undisturbed in the deep IR and, if this was the case, what type of behavior it follows.

\section{Numerical Method}\label{sec:num_meth}

The results displayed in the previous section suggest that the solution to the TBA equation \eqref{tbas}, for S-matrices of the form \eqref{sNpole}, may generally possesses a singular dependence on the parameter $R$. In particular the slope of the tangent to the graph of $E(R)$ apparently diverges at some $R=R_*$. Such critical points are known as \emph{turning points}. Their presence in the dependence of the ground-state energy $E$ on the system size $R$ evokes the case of the TTbar deformed models, in which all the quantities obtainable from the TBA display a square-root singularity at the same value $R=R_*$.

The iterative procedure described in \S \ref{subsec:iterative} becomes unstable at $R\to R_*$, therefore it is not particularly suitable for analyzing the vicinity of the singular point. Fortunately, 
many powerful methods exist that are capable of handling numerically critical points in non-linear equations. We refer to the nice monograph by Allgower and Georg \cite{allgower2012numerical} for an introduction, paired with an extensive literature, to the subject. The simplest of these numerical routines is the already mentioned PALC method which, in spite of the simplicity of its implementation, will be entirely sufficient to handle the situations of interest for us. In this section we will quickly review this method and its main features.

\subsection{The pseudo-arc-length continuation method}
Before starting let us point out a trivial fact: the TBA equation \eqref{tbas} is \emph{non-linear}. It is then not at all surprising that its solutions can develop a highly non-trivial dependence on the parameters. Conversely, what is remarkable is that in the vast majority of instances known in the literature, the solution to the TBA equations display a simple behavior as functions of $R$. In full generality, we should expect a solution $\epsilon(\theta\vert R)$ to potentially present, as a function of $R$\footnote{In principle, the solution might possess critical points also in its dependence on the other parameters present in the TBA equation. We found no hint of such a possibility and we will thus simplify our discussion by concentrating on the dependence on the parameter $R$.}, any type of critical point imaginable. As we will see later, in the cases of the 1CDD and 2CDD models we are concerned with here, only turning points appear. We will thus restrict our attention to the simple cases in which every critical point is a turning point. This considerably simplifies both the discussion and the actual implementation of the PALC method, although, if needed, it is entirely possible -- and not exceedingly difficult -- to include the existence of bifurcations in the game.

Since our goal is to analyze the TBA equation \eqref{tbas} numerically, we are going to describe the principles of the PALC for maps between finite-dimensional spaces. Let us then truncate and discretize the real $\theta$-line on a $N$-point lattice $\left\lbrace\theta_k\;\vert\; k=1,2,\ldots ,N\right\rbrace$ which, for the moment, we are not going to specify further. Now, consider a parametrized map $H$ which takes as input a parameter $R\in\mathbb R$ together with the values $\epsilon_k=\epsilon(\theta_k)\in\mathbb R$ of some real function on the lattice, and yields $N$ real numbers:
\begin{eqnarray}
	H\;:\quad \begin{array}{c c c}
		\mathbb R^{N}\times \mathbb R & \longrightarrow & \mathbb R^{N}\\
		 &  & \\
		(\vec{\epsilon},R) & \longmapsto & \vec{H}(\vec{\epsilon},R)
	\end{array}\;,
\end{eqnarray}
where we packaged the values $\epsilon_k$ and $H_k$ into vectors $\vec{\epsilon}$ and $\vec{H}$. We wish to explore the following fixed-point condition
\begin{eqnarray}
	\vec{H}(\vec{\epsilon},R) = \vec{0}\;.
\label{eq:map_equation}
\end{eqnarray}
Note that the TBA equation \eqref{tbas}, appropriately discretized and truncated, can be written in the above form. By definition, the map $H$ acts between spaces of different dimensionality, meaning
\begin{eqnarray}
	\textrm{dim}[\textrm{Ker}(H)]\geq 1\;,
\end{eqnarray}
or, in other words, the image of the null vector $\vec{0}\in\mathbb R^N$ under the inverse map $H^{-1}$ is a space of dimension at least $1$. Hence at a generic point, where $\textrm{dim}[\textrm{Ker}(H)] = 1$, this image is a curve
\begin{eqnarray}
	C\;:\quad J\subset \mathbb R\;\longrightarrow \; \mathbb R^N\times \mathbb R\;.
\end{eqnarray}
We call this the \emph{solution curve} for the map $H$.

Our goal is to follow the solution curve from a given starting point $C_i = (\vec{\epsilon}_i, R_i)$ to a final one $C_f = (\vec{\epsilon}_f, R_f)$. The most straightforward way to achieve this is to simply parametrize the curve by $R$ and employ some numerical iterative routine, such as the one reviewed in \S \ref{subsec:iterative}, to move from $C_i = C(R_i)$ to $C_f = C(R_f)$. However this simple-minded approach fails at any point in the parameter space where the rank of the Jacobian
\begin{eqnarray}
	\mathcal{J}_{kl} = \frac{\partial H_k}{\partial \epsilon_l}\;,
\end{eqnarray}
is not maximal. There we can no longer rely on the implicit function theorem to solve (\ref{eq:map_equation}) for $\vec{\epsilon}$ in terms of $R$. More geometrically, what happens is that the curve $C(R)$ displays a turning point, where $\frac{d}{dR}C(R)$ diverges. Fortunately there exists a very simple cure for this problem: instead of parameterizing the curve $C$ by the parameter $R$, we can use an auxiliary quantity $s$, traditionally chosen to be the arc-length of $C$ or a suitable numerical equivalent, whence the name \emph{pseudo-arc-length} given to this approach. The condition (\ref{eq:map_equation}) then becomes
\begin{eqnarray}
	\vec{H}(C(s)) = \vec{0}\;,\qquad s\in J\subset\mathbb R\;.
\label{eq:eq_map_H}
\end{eqnarray}
In order to proceed, let us take a derivative of this condition with respect to the parameter $s$. We immediately obtain
\begin{eqnarray}
	H'(C(s)) \dot{C}(s) = \vec{0}\;,
\end{eqnarray}
where the \emph{extended Jacobian}
\begin{eqnarray}
	H'(C(s)) = \Bigg(\;\mathcal J\;\Bigg\vert\;\frac{d\vec{H}}{dR}\;\Bigg)\;,
\end{eqnarray}
is a $N\times(N+1)$ block matrix, while
\begin{eqnarray}
	\dot{C}(s) = \left(\begin{array}{c}
		\frac{d}{ds} \vec{\epsilon} \\[0.1cm]
		\hline\\[-0.4cm] \frac{d}{ds} R
	\end{array}\right)\;,
\end{eqnarray}
is an $(N+1)$ column vector. At this point we seem to be short of $1$ condition, since we introduced an additional parameter. However, remember that we decided to choose $s$ as the (pseudo-)arc-length of $C$, which means
\begin{eqnarray}
	\vert\vert \dot{C}(s)\vert\vert = 1\;.
\end{eqnarray}
Summing up, we converted our non-linear problem, supported by the starting point $(\vec{\epsilon}_i,R_i)$, into an initial value problem
\begin{eqnarray}
	H'(C(s))\dot{C}(s) = \vec{0}\;,\qquad \vert\vert\dot{C}(s)\vert\vert = 1\;,\qquad C(s_i) = (\vec{\epsilon}_i,R_i)\;,
\label{eq:in_val_prob_map_H}
\end{eqnarray}
capable of dealing with the presence of turning points. Still, this formulation is somewhat unnatural as it completely disregards the fact that the curve $C$ is the fixed point of the map $H$, and, as such, should enjoy powerful local contractive properties with respect to iterative solution methods -- such as Newton's method. We are then led to an integrated approach in which we numerically integrate (\ref{eq:in_val_prob_map_H}) very coarsely and subsequently employ some kind of iterative method to solve (\ref{eq:eq_map_H}) locally. This is the general strategy behind the approaches known as \emph{predictor-corrector routines}. In Appendix \ref{app:pred_corr} we are going to describe the one that we employed in this work and present a pseudo-code of its implementation.

\section{Results for the 2CDD model}\label{sec:results}

Here we present some results obtained using the numerical techniques of the previous Section.
We first concentrate on the fermionic 2CDD models and then discuss some facts about the bosonic models.

\subsection{Fermionic case}

The numerical data we collected, of which we have shown some example in \S \ref{subsec:two_branches}, strongly indicate the following properties of the ground-state energy $E(R)$ as a function of $R$:
\begin{itemize}
	\item[--] $E(R)$ is a double-valued function of $R$, in the range $R>R_{\ast}$ with values in the negative real numbers;
	\item[--] The point $R=R_{\ast}$ is a square-root branching point -- or, using the terminology of \S \ref{sec:num_meth}, a turning point -- of the function $E(R)$;
	\item[--] There is no sign of additional turning or singular points other than $R=R_{\ast}$;
	\item[--] The two branches display the large-$R$ behaviors \eqref{eq:E_asymp_primary} and \eqref{eq:E_asymp_secondary}.
\end{itemize}
We could not find a convincing analytic argument proving the first three properties and we regard them as experimental observations. On the other hand, the last property \eqref{eq:E_asymp_secondary} can be verified analytically, as we are now going to show.

\subsubsection{The large $R$ behavior}\label{subsec:large_R_fermion}

Let us analyze the possible behaviors of the TBA equation \eqref{tbas} at large $R$. To this end, we write the equation as follows
\begin{eqnarray}
	\epsilon(\theta) = d(\theta) -\chi(\theta)\;,
\label{eq:TBA_symbolic}
\end{eqnarray}
where $d(\theta)$ is the driving term and $\chi(\theta)$ the convolution:
\begin{eqnarray}
	d(\theta) = R\cosh\theta\;,\qquad \chi(\theta) =  \int\,\varphi(\theta-\theta')\,\log\left[1+e^{-\epsilon(\theta')}\right]\,\frac{d\theta'}{2\pi}\;.
\end{eqnarray}
As $R\rightarrow \infty$, the driving term becomes large, $\sim R$, and, in order for the equation \eqref{eq:TBA_symbolic} to be satisfied, it has to be balanced by a similar behavior in either $\epsilon(\theta)$, $\chi(\theta)$ or both. The standard assumption is that
\begin{eqnarray}
	\epsilon(\theta)\underset{R\rightarrow\infty}{\sim} d(\theta)\;,\qquad \chi(\theta)\underset{R\rightarrow\infty}{\ll}d(\theta)\;,
\label{eq:standard_large_R}
\end{eqnarray}
which turns out to be consistent, since, as one easily verifies,
\begin{eqnarray}
	\chi(\theta) \underset{R\rightarrow\infty}{\sim} \int\,\varphi(\theta-\theta')\,\log\left[1+e^{-R\cosh\theta'}\right]\,\frac{d\theta'}{2\pi} \underset{R\rightarrow\infty}{\sim} \frac{\varphi(\theta)}{\sqrt{2\pi R}} e^{-R} \underset{R\rightarrow\infty}{\ll} R \cosh\theta\;.
\label{eq:standard_large_R_chi}
\end{eqnarray}
However this is not, in general, the only possibility. It might be the case that the convolution term $\chi(\theta)$ is diverging as $R\rightarrow \infty$ and becomes comparable with either $\epsilon(\theta)$, $d(\theta)$ or both. It is then not difficult to check that only two possibilities are consistent:
\begin{enumerate}
	\item $\epsilon(\theta) \underset{R\rightarrow\infty}{\longrightarrow} 0$ \underline{and} the kernel $\varphi(\theta)$ is not integrable on the real line;
	\item $\epsilon(\theta) \underset{R\rightarrow\infty}{\sim} -R\,f(\theta)$ where $f(\theta)$ is positive only in some finite\footnote{The subset $\Theta$ cannot be infinite, since the equation \eqref{eq:TBA_symbolic} forces $\epsilon(\theta)$ to behave as $d(\theta)$ for $\theta\rightarrow \pm \infty$.} subset $\Theta\subset\mathbb{R}$ of the real line and negative everywhere else.
\end{enumerate}
The scenario 1 cannot arise for the class of models we are dealing with\footnote{This scenario is, however, possible in models whose $S$-matrix presents a non-vanishing factor $\Phi_{\textrm{entire}}(\theta)$ \eqref{phientire}. In particular it describes the large $R$ behavior of the secondary branch $\tilde{E}(R)$ in the $T\bar{T}$-deformed theories.}, since the kernels \eqref{eq:TBA_kernel_v2} are obviously bounded functions of $\theta\in\mathbb{R}$. The situation 2 is, on the other hand, a possible one. Let us explore its consequences.

In the hypothesis that
\begin{eqnarray}\label{eq:second_large_R}
	\epsilon(\theta) \underset{R\rightarrow\infty}{\sim} -R\,f(\theta)\;,\qquad \left\lbrace\begin{array}{l l} f(\theta) > 0\;,& \theta\in\Theta\subset\mathbb{R}\;, \\ f(\theta)\leq 0& \theta\in\Theta^{\perp} = \mathbb{R}-\Theta \;,\end{array}\right.
\label{eq:negative_epsilon}
\end{eqnarray}
the convolution can be approximated as follows
\begin{eqnarray}
	\chi(\theta) \underset{R\rightarrow\infty}{\sim} R\intop_{\Theta}\,\varphi(\theta-\theta')\,f(\theta')\,\frac{d\theta'}{2\pi} + \intop_{\mathbb{R}}\,\varphi(\theta-\theta')\,\log\left[1+e^{-R\left\vert f(\theta')\right\vert}\right]\,\frac{d\theta'}{2\pi}\;.
\end{eqnarray}
Discarding the second term in the right-hand side, we arrive at the linear equation
\begin{eqnarray}
	f(\theta) = -\cosh\theta + \intop_{\Theta}\,\varphi(\theta-\theta')\,f(\theta')\,\frac{d\theta'}{2\pi}\;.
\end{eqnarray}
Due to our hypothesis on the function $f(\theta)$, we see that the integrand in the right-hand side above is positive for any $(\theta,\theta')\in\mathbb R^2$, which implies the following bound
\begin{eqnarray}
	0\leq \intop_{\Theta}\,\varphi(\theta-\theta')\,f(\theta')\,\frac{d\theta'}{2\pi} \leq \underset{t\in\Theta}{\textrm{Max}}\left[f(t)\right]  \intop_{\Theta}\,\varphi(\theta-\theta')\,\frac{d\theta'}{2\pi} \;.
\end{eqnarray}
Now, let $\theta_{\textrm{M}}\in\Theta$ be such that $f(\theta_{\textrm{M}}) = \underset{t\in\Theta}{\textrm{Max}}\left[f(t)\right]$, then the following inequalities are true
\begin{eqnarray}
	-\cosh\theta_{\textrm{M}} \leq f(\theta_{\textrm{M}}) \leq -\cosh\theta_{\textrm{M}} +f(\theta_{\textrm{M}}) \intop_{\Theta}\,\varphi(\theta_{\textrm{M}}-\theta')\,\frac{d\theta'}{2\pi} \;.
\end{eqnarray}
Rearranging the right inequality above, we find that
\begin{eqnarray}
	\intop_{\Theta}\,\varphi(\theta_{\textrm{M}}-\theta')\,\frac{d\theta'}{2\pi} \geq 1+\frac{\cosh\theta_{\textrm{M}}}{f(\theta_{\textrm{M}})} > 1\;,
\end{eqnarray}
which we can interpret as a constraint on the class of models which allow for this scenario. In fact, remember that the integral of the kernel on the whole real line, \eqref{eq:L1_kernel}, counts the number $N$ of CDD factors appearing in the $S$-matrix \eqref{sNpole}. But, since we assumed that $\Theta$ is a finite subset of $\mathbb R$, we find that
\begin{eqnarray}
	N > \intop_{\Theta}\,\varphi(\theta_{\textrm{M}}-\theta')\,\frac{d\theta'}{2\pi} > 1\quad \Longrightarrow \quad N>1\;.
\label{eq:bound_on_N}
\end{eqnarray}

Thus we have found that the fermionic 1CDD models, namely sinh-Gordon and the staicase models, can only display the standard large $R$ behavior \eqref{eq:standard_large_R}, \eqref{eq:standard_large_R_chi}. We stress that this result should not be read as a proof of the absence of turning points in these models, but rather as a sanity check for the correctness of our computations, since the ground-state energy for fermionic 1CDD models is well known to be a smooth and monotonously increasing function of the radius in the whole range $R>0$. Conversely, all fermionic $N$CDD models with $N>1$ allow for both the standard large $R$ behavior \eqref{eq:standard_large_R}, \eqref{eq:standard_large_R_chi} and the non-standard one \eqref{eq:negative_epsilon}. Consequently, their ground-state energy will possibly display both the asymptotic behavior \eqref{eq:E_asymp_primary} and \eqref{eq:E_asymp_secondary}, where
\begin{eqnarray}
	\varepsilon_{-} = \intop_{\Theta}\,\cosh\theta\,f(\theta)\,d\theta\;,
\end{eqnarray}
in accordance with the numerical data we have obtained.

\subsubsection{Analysis of the numerical data}\label{sec:F2CDD_num}

\begin{figure}[t!]
\begin{center}
\includegraphics{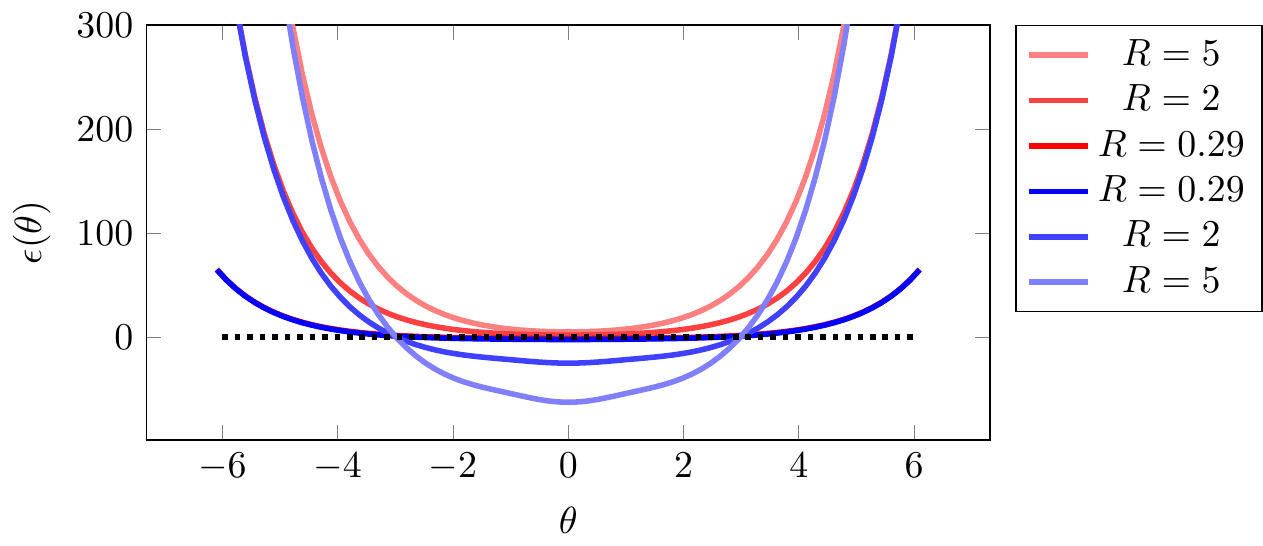}
\end{center}
\caption{
Pseudoenergy $\epsilon(\theta)$ for the secondary branch solution (blue) at large values of $R$, showing the expected behavior \eqref{eq:second_large_R}, namely it is below $0$ (marked with the dashed line) in a finite interval. Corresponding behavior of the iterative solution (red). Here the model parameters are $\theta_0=2$ and $\gamma=4\pi/10$, though we checked the qualitative picture to remain the same within the whole set of admissible values of $\theta_0$ and $\gamma$.}
\label{fig:second_branch_large_R}
\end{figure}

The fermionic 2CDD models were classified in \S \ref{sec:models} into cases (a) to (d). We have performed numerical analysis for all the different cases and the results show that the behaviors are qualitatively the same. Thus, we are going to show here the details of the numerical analysis only for the representative case (d).  
We begin by analyzing the numerical solution obtained through the PALC method for large values of $R$. It was argued in the previous section that the pseudoenergy should behave as in \eqref{eq:second_large_R}, assuming negative values in a finite subset of the real line and positive values elsewhere. This is indeed checked to be true for all the 2CDD models under consideration, as illustrated for a particular member of this family in Figure \ref{fig:second_branch_large_R}, and to be contrasted with the standard iterative solution (the primary branch) which is positive everywhere. The numerics indicate that the negativity region is always a single interval centered at the origin of the form $\Theta=\{\theta\in\mathbb{R}\,|\,-\Lambda\le\theta\le\Lambda\}$. They also indicate that the interval size $\Lambda$ is model-dependent. In particular, it seems to grow with $\theta_0$ and decreases with $\gamma$. Nevertheless, the precise dependence of $\Lambda$ on the parameters deserves further investigation.   

\begin{figure}[t!]
\begin{center}
\includegraphics{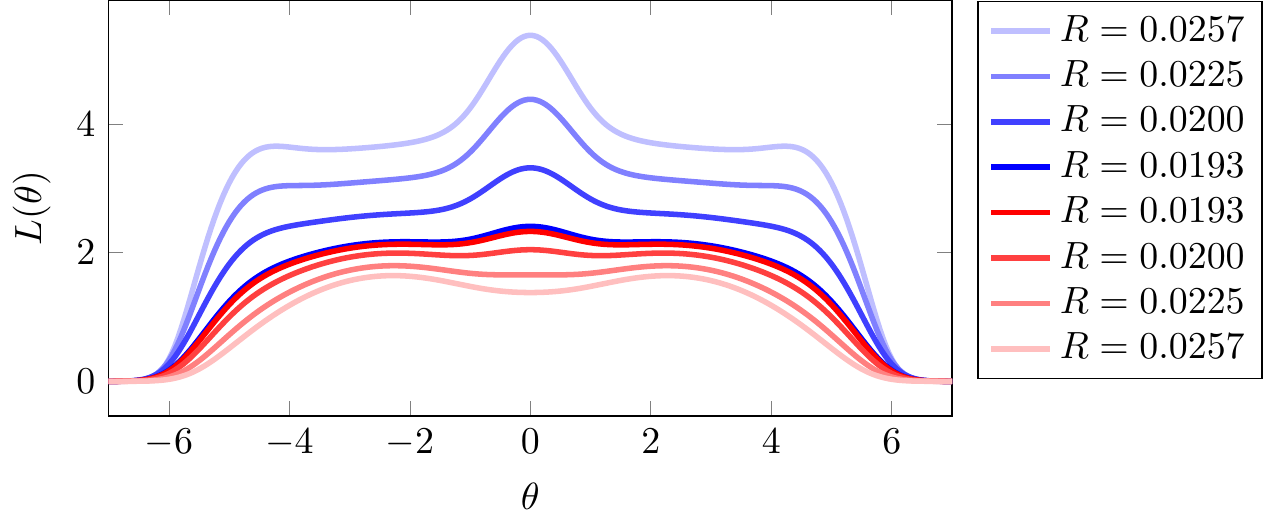}
\end{center}
\caption{$L(\theta)$ for both the primary (red) and secondary (blue) branch solutions as $R$ approaches the critical value $R_*$. For each color (blue or red), the color gradient indicates the decrease of $R$ towards $R_*$, where the two branches merge. Here $\theta_0=5$ and $\gamma=4\pi/10$, which lead to $R_*\approx0.0192$. 
}
\label{fig:both_branches_R_crit}
\end{figure}

We then proceed to analyze the secondary branch solution in the opposite extremum of $R$, i.e., as $R$ approaches the critical value $R_*$. For some of the plots it will be convenient to show the results in terms of the log-scale distance
\begin{equation}
    x=\log(R/2)
\end{equation}
that alleviates the exponential dependence (with $x_*=\log(R_*/2)
$ for the corresponding critical point). Here we find it more instructive to display $L(\theta)$ instead of the pseudoenergy itself in order to ease the comparison with the primary branch solution. The situation is illustrated in Figure \ref{fig:both_branches_R_crit}. The two branches approach each other as the value of $R$ decreases, eventually merging at $R=R_*$ after which they become complex-valued. For each $R$, the function $L(\theta)$ for the secondary branch is everywhere larger than the corresponding primary branch counterpart, which is compatible with the previously mentioned fact that it has lower energy (recall the overall minus sign in \eqref{etbas}).

The critical value $R_*$ could in principle have a dependence on $\theta$. We ran an extensive numerical test exploring this possibility, but all the numerical results indicate $\theta$-independence to high accuracy, even though at this moment we do not have an 
analytic proof of this property.   
The analyses were as follows. 
We first ran the iterative numerical routine
and computed the pseudoenergy $\epsilon(\theta)$ for at least 
ten different values of $x$ differing 
from each other and from $x_*$ by $10^{-8}$.
Then, we selected several values of $\theta$
and for each value we performed a square root fit of the form $a(\theta) + b(\theta) \sqrt{-x_*(\theta) + x}$. 
The fits were done using Mathematica's \texttt{NonlinearModelFit} function by giving 
an initial estimate for $x_*(\theta)$. 
By comparing all the obtained $x_*(\theta)$, we verified that they agree up to errors greater than $10^{-8}$ which was our minimal working  precision.
The analysis was performed for several values of $\theta_0$ and for $\gamma$ in the range $0 \le \gamma \le (99/200) \pi$. In many cases, when the number of
necessary points in the discretized $\theta$ grid was not very high it was possible to work with even higher precision. 
In those cases, another way of 
getting 
$x_{*}$ with high precision 
is by assuming a square root 
behavior for the pseudoenergy
and solving the resulting equations 
using Mathematica's \texttt{FindRoot} function. 

\begin{figure}[t!]
    \centering
     \begin{subfigure}[b]{6cm}
         \centering
         \includegraphics[width=\textwidth]{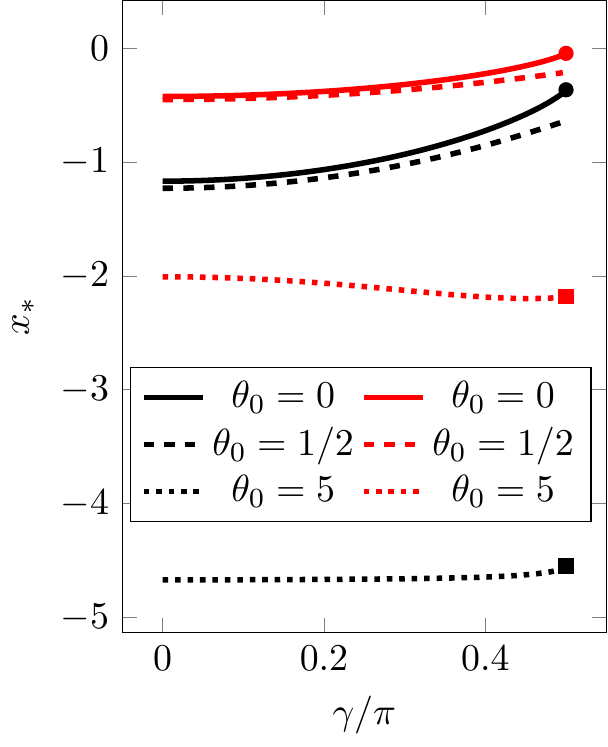}
         \caption{$\gamma$ dependence of $x_*$.}
         \label{fig:xcgam}
     \end{subfigure}
     \begin{subfigure}[b]{6.31cm}
         \centering
         \includegraphics[width=\textwidth]{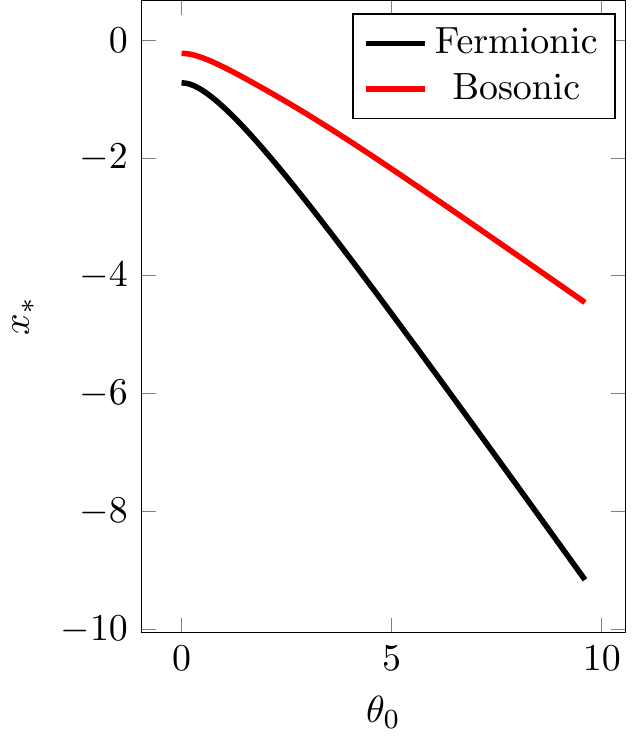}
         \caption{$\theta_0$ dependence of $x_*$ with $\gamma=2\pi/5$}
         \label{fig:xcth0}
     \end{subfigure}

    \caption{Dependence of the critical $x_*$ in the model parameters. Black lines correspond to fermionic 2CDD models, red lines correspond to bosonic ones. On (a), we demonstrate the validity of the narrow resonance limit approximation for $x_*$ (red and black bullets/boxes), see in \ref{sec:NRlimit}. }
    \label{fig:xcdeps}
\end{figure}

In addition, we also verified that $R_*$ depends smoothly on the model parameters $\theta_0$ and $\gamma$, as shown in Figure \ref{fig:xcdeps} for both the fermionic and bosonic models. In particular, for large $\theta_0$ we have the asymptotic behavior $x_*=\log(R_*/2)\approx-\theta_0+x_*^{(0)}$ (see \S\ref{sec:NRlimit} for a derivation in the special limit where $\gamma$ is close to $\pi/2$, for which $x_*^{(0)}=\log\log(2+2\sqrt{2})$; for other values of $\gamma$ the linear term remains the same, though $x_*^{(0)}$ is different).

\subsection{Bosonic case}
We have also repeated the analysis described above using the PALC method to the case of bosonic systems. The numerical routine used in this case only differs from the fermionic case by a few signs. As already mentioned in \S\ref{sec:models}, the solutions to the TBA equation for the bosonic models have intricate behavior already for the 1CDD cases. It was first noticed in 
\cite{Mussardo:1999aj} (for the case of real 
$u_1$, in the notation of \eqref{1cdd}) that the numerical iterative routine stops converging for some $R_{*}$, signaling the presence of a singularity. In fact, we have verified numerically that all the bosonic models up to two CDD factors behave
similarly to the fermionic 2CDD models of previous section, i.e.,
they have a ``primary branch” and a
``secondary  branch” which merge at a critical scale $R_{*}$, where the energies $E(R)$ have square-root 
singularities in $R_{*}$, and the value of $R_{*}$ is independent of $\theta$. 

There is a simple argument based on the well-known relation between bosonic and fermionic TBA which makes this behavior of the bosonic $1$CDD model rather natural. Consider the TBA equation \eqref{tbas}, \eqref{Ldef} with $\sigma = +1$ and an $N$CDD kernel \eqref{sNpole}, and introduce the following function 
\begin{eqnarray}
    \tilde{\epsilon}(\theta) = \log\left[e^{\epsilon(\theta)} - 1\right]\;.
\label{eq:epsilon_tilde}
\end{eqnarray}
Some trivial manipulations show that this function satisfies a fermionic TBA equation with kernel
\begin{eqnarray}
    \tilde{\varphi}(\theta) = \varphi(\theta) + 2\pi \delta(\theta)\;,
    \label{eq:BosonicFermionic} 
\end{eqnarray}
with the $\delta(\theta)$ being the Dirac $\delta$-function. Therefore, a general bosonic $N$CDD model is equivalent to the ($N+1$)CDD fermionic TBA, taken in the limit when $u_{N+1}\to 0$ (see \eqref{eq:kernelNCDD})\footnote{Notice that $\lim_{u\to 0}\text{log}\frac{i\sin u - \sinh\theta}{i\sin u - \sinh\theta} = i\pi\,\text{sign}(\theta)$, for the principal branch of the log function.}. Recalling the
arguments presented in \S \ref{subsec:large_R_fermion}, we conclude that bosonic $N$CDD models admit two different types of large $R$ behaviors whenever $N>0$.

The large $R$ regime of the pseudoenergy $\epsilon(\theta)$ for the primary branch 
is as expected and it is easily accessed  numerically, however for  
the secondary branch it is more 
involved to compute it.
By increasing the value of $R$, eventually we reach a value $R^{\prime}$ where the PALC method suddenly ceases to provide a real solution and reverts back to the primary branch solution. 
Analyzing the behavior
of $\epsilon(\theta)$ for complex values of $\theta$,
we verified that a pair of complex conjugate zeros of $z(\theta)=1-e^{-\epsilon(\theta)}$ is approaching the real axis and causing the numerical instability. In principle it is possible to refine the numerical methods so as to obtain solutions for $R>R^{\prime}$. 
However, it is not clear at the moment whether or not
those singularities of $L(\theta)$ ever cross the real axis. In case they do, an analysis similar to the one performed in \cite{Dorey:1996re} for the excited state TBA could be carried out. 
We leave the analysis of the large $R$ behavior of the secondary branch in bosonic models for a future study.

\begin{figure}[t!]
\begin{center}
\includegraphics{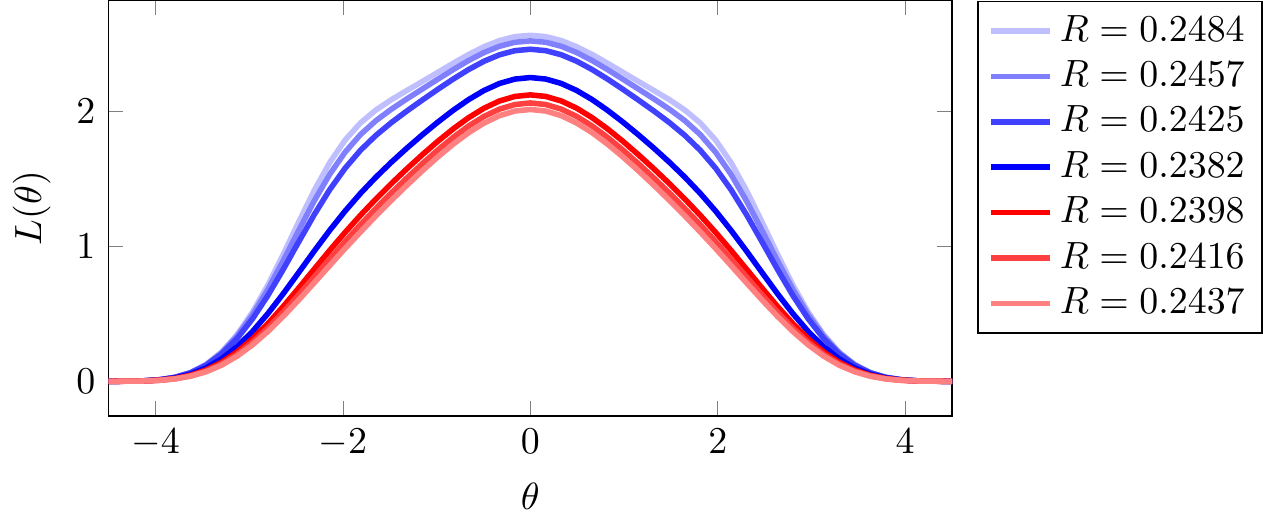}
\end{center}
\caption{$L(\theta)$ for
the 2CDD bosonic model of type (d) with  
$\theta_0=5$ and $\gamma=3\pi/10$, in which case $R_*\approx0.2382$.
Similarly to the fermionic case the function $L(\theta)$ for the secondary branch is everywhere greater than the one for the first branch.}
\label{1CDDbosonicoL}
\end{figure}

The behavior of the models for 
$R$ close to $R_{*}$ is illustrated
in Figure 
\ref{1CDDbosonicoL} by the $L(\theta)$ function for a 2CDD model of type (d). 
The qualitative picture is similar to the fermionic case, i.e. the function $L(\theta)$ for the secondary branch solution is greater everywhere than the one for the primary branch and the two merge as the critical point is approached.
We conclude this subsection by showing in 
Figure \ref{fig:xcdeps}  the smooth dependence of $x_{*}$ on the model parameters and in particular in the limit $\gamma \rightarrow \pi/2$. 
{In addition, notice that the bosonic curve is always 
above of the fermionic curve for the 
same parameters. 
This can be understood by analyzing the map \eqref{eq:BosonicFermionic} and the fact
that the additional delta function term 
always give a positive contribution to the
convolution term of the TBA equations.}

\subsection{Narrow resonance limit}
\label{sec:NRlimit}
Here we consider the special limit $\gamma\to\frac{\pi}{2}$ of the kernel~\eqref{eq:2cdd_kernel}.
In this limit the poles of the kernel get closer to the real line, finally forming two Dirac $\delta$ functions. We shall refer to this as the Narrow Resonance (NR) limit. After integration of the delta functions and exponentiation, TBA \eqref{tbas} becomes the difference equation
\begin{equation}
\label{eq:NR}
 Y(\theta|R)= e^{-R\cosh \theta}[1-\sigma Y(\theta+\theta_0|R)]^{-\sigma}[1-\sigma Y(\theta-\theta_0|R)]^{-\sigma}\,,
\end{equation}
where we introduced the notation $Y(\theta|R)=e^{-\epsilon(\theta|R)}$. Note that this can be seen as an infinite set of equations relating the values of $Y$ on the grid points $\theta \in (-\theta_0,\theta_0)+\theta_0\mathbb{Z}$.

Let us focus on the fermionic case ($\sigma=-1$). Introducing $y_k=Y(\theta+k \theta_0)$ and $g_k=e^{-R \cosh(\theta+k \theta_0)}$ we can write \eqref{eq:NR} as
\begin{equation}
    y_k=g_k(1+y_{k-1})(1+y_{k+1}) \, , \qquad (k\in\mathbb{Z})
\end{equation}
and look for a solution for different grids specified by a choice of $\theta$. This is an infinite set of equations, however starting with $k=0$ one can obtain an approximate solution by truncating the system for some $|k|\leq m$, since $g_k$ and $y_k$ decay very rapidly with $R$ and $\theta_0$, and hence with $k$.
Truncating to $m=1$ leads to the quadratic equation
\begin{equation}
    \label{eq:NRtr1}
    y_0=k_0[1+g_1(y_0+1)][1+g_{-1}(y_0+1)]\,.
\end{equation}
One can now choose the integer lattice (i.e., $\theta=0$), to get
\begin{align}\label{eq:y0nrm}
 y_0 &=-1-e^{-R\cosh\theta_0}+\frac{1}{2}e^{R(1+2\cosh \theta_0)}\left(1\pm\sqrt{1-4e^{-R(1+\cosh\theta_0)}(1+e^{-R\cosh \theta_0})}\right).
\end{align}
The solution develops a square root singularity at $x_*\approx-\theta_0+\log\log(2(1+\sqrt{2}))$, which is compatible with our findings in \S\ref{sec:F2CDD_num}. This point is shown as a red bullet in Figure \ref{fig:xcgam}. In contrast to the general case, it is clear that here the branching point depends on the choice of $\theta$ lattice. Let us also comment that a similar analysis of the truncated system in the bosonic case ($\sigma=+1$) using the half-integer lattice leads to the black box shown in Figure \ref{fig:xcgam}.\footnote{ The analog of \eqref{eq:y0nrm} comes with a more complicated square root argument and no analytical solution for $x_*$ as a function of $\theta_0$ can be found in that case, although it is straightforward to find it numerically.}

Note that the truncation to $m=1$ is only valid for sufficiently large $R$ and $\theta_0$. Increasing the truncation order leads to more coupled equations, which in turn can be recast as an (more complicated) algebraic equation for $y_0$, with parameters depending on $\theta$. The number of solutions increases accordingly. However, for any $\theta \in (-\theta_0,\theta_0)$ there is always a single pair of solutions which collide and form a  branching point at some $x_*(\theta)\approx -\theta_0 + \text{const.}$, corresponding to real, positive $R_*(\theta)$, a feature that is not altered by increasing the truncation order.

Finally, we remark that in the further special limit $\theta_0=0$, the difference equations~\eqref{eq:NR} become simple algebraic equations for $Y(\theta)$ that can be exactly solved both in the fermionic and the bosonic case, leading to exact expressions $x_*=\log \log 2$ and $x_*=\log \log \frac{3}{2}\sqrt{3}$ for the critical points, respectively. These points are also shown in Figure~\ref{fig:xcgam}, emphasizing the smooth nature of the limit $\gamma\to\frac{\pi}{2}$.

\begin{figure}
    \centering
    \includegraphics{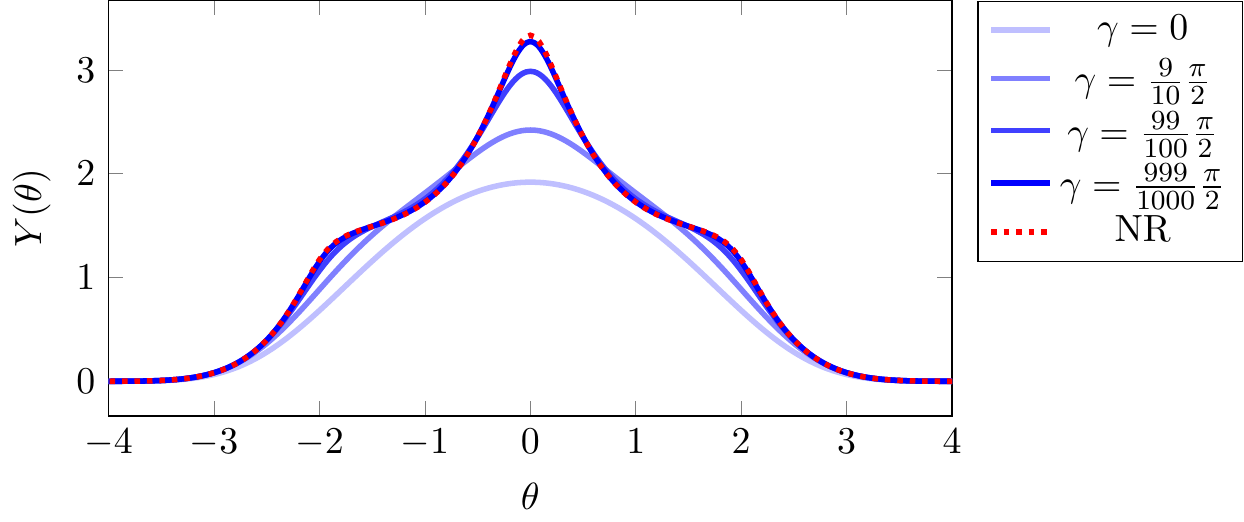}
    \caption{Approaching the Narrow Resonance (NR) limit for $\theta_0=2$ and $x=1.75$}
    \label{fig:NRlimit}
\end{figure}

In Figure~\ref{fig:NRlimit} we present as an example a solution with $m=8$ truncation together with the iterative solution of the integral equation \eqref{tbas} for $\theta_0=2$ and $\gamma$ approaching $\pi/2$, just before reaching the (first) critical $R_*(\theta)$ of the NR limit. The transition seems to be smooth, however we do not yet have a complete understanding of the nature of this limit. We plan to revisit the narrow resonance model in a more sistematic way in the future.

\section{Discussion}\label{sec:discussion}

There are two general questions which we believe our results shed some light upon. One concerns the short-distance behavior of the theory under the generalized TTbar deformation
\eqref{Aalphas}. Our results supports the expectation that, at least in the cases when the CDD factor in the associated $S$-matrix deformation has the form \eqref{cddn}, \eqref{phipole} with finite $N$, the theory develops the Hagedorn singularity corresponding to a density of high-energy states much greater than what is allowed in a Wilsonian QFT. Although we demonstrated this in a limited set of examples -- the 2CDD deformations of the free $S$-matrix with both fermionic and bosonic statistics and the 1CDD deformations of the free boson $S$-matrix -- this result likely extends to more general $N$CDD deformations, at least for massive theories involving only one kind of particles. In fact the case $N=\infty$, a model known as \emph{Elliptic sinh-Gordon}, is shown to display the same behavior as the ones studied here \cite{Cordova:2021fnr}. We note that this behavior is qualitatively the same as the one encountered under the ``TTbar proper'' deformation \eqref{Aalpha} of a generic local QFT. Moreover, the singularity of $E(R)$ at the Hagedorn point $R_*$ is a square-root branching point, exactly as in the TTbar deformations with negative $\alpha$. From a formal point of view, this nature of the singularity is not entirely unexpected. Indeed, the character of the singularity relates to the rate of approach of the Hagedorn asymptotic \eqref{hagedorn1} at high energy $\mathcal{E} \to \infty$. Assume that the approach is power-like\footnote{{It is interesting to compare this assumption with the analysis of thermodynamic stability in \cite{Barbon:2020amo}.}}
\begin{eqnarray}\label{tohagedorn}
S(\mathcal{E}) = R_*\,\mathcal{E} - \frac{a\,L^{\kappa+1}}{\mathcal{E}^\kappa} + \cdots
\end{eqnarray}
where $\kappa$ is some positive number, $L$ is the spatial size of the system which is assumed to be asymptotically large, and the dots represent yet higher negative powers of $\mathcal{E}$. The dependence on $L$ of the subleading term reflects the extensive nature of the entropy, which must behave as $L\,\sigma(\mathcal{E}/L)$ in the limit $L\to\infty$, with the intensive quantity - the entropy density $\sigma$ - depending on the energy density $\mathcal{E}/L$. Inspection of \eqref{tohagedorn} reveals the mass dimension of the coefficient $a$ to be $a\sim[\text{mass}]^{2\kappa+1}$. Having in mind that all the deformation parameters $\alpha_s$ in \eqref{Aalphas} have even integer dimensions, one could expect that the exponent $2\kappa+1$ is an integer. The lowest positive $\kappa$ consistent with this assumption is $\kappa=1$, and then \eqref{tohagedorn} leads exactly to the square-root singularity of $E(R)$. Still, the physics behind this simple character of the singularity appears mysterious. Analytic continuation of $E(R)$ below $R_*$ returns complex values of $E$. This likely signals an instability of the ground state at $R<R_*$ against some sort of decay. If so, what is the product(s) of the decay? Usually in a theory with finite range of interaction the decay of the unstable ground state goes through the process of nucleation, as in the ``false vacuum'' decay studied in \cite{Coleman:1977py,Kobzarev:1974cp}. However such a decay would imply a much weaker -- and analytically more complicated -- singularity at $R_*$. Therefore the simple algebraic character of the actual singularity appears puzzling. A different, but possibly related, question is the physical interpretation of the secondary branch of $E(R)$ discovered in \S \ref{sec:results}.

An even more general question concerns the relation between the $S$-matrix and the underlying local structure. Suppose we are given an $S$-matrix, i.e. a collection of masses of stable particles as well as the full set of scattering amplitudes, satisfying all the standard requirements of the $S$-matrix theory - unitarity, analyticity, crossing and bootstrap conditions (see e.g. \cite{Eden:1966dnq, Iagolnitzer:1994xv}), with the singularity structure consistent with the macro-causality \cite{Iagolnitzer:1974wz}. Is there a local QFT generating such a scattering theory? The answer is generally no. There are consistent $S$-matrices which cannot be derived from Wilsonian QFT, and indeed do not have an underlying local structure, meaning a complete algebra of local operators. This possibility is famously realized in string theories. The results presented here support the expectation that the overwhelming majority of self-consistent $S$-matrices are not derivable from local QFT. Although this expectation arise from a general analysis of the RG flows \cite{Wilson:1973jj}, we substantiate it by providing concrete examples in 1+1 dimensions with factorizable $S$-matrices consisting of pure CDD factors. We studied a number of examples of such $S$-matrices and verified that they lead to the Hagedorn density of high-energy states \eqref{hagedorn1}, familiar to the string theories. What's more, it looks likely that this situation is rather general: with the exception of a small subset of ``local field-theoretic'' $S$-matrices, the bulk part of the space of consistent, factorizable S-matrices in 1+1 dimensions, leads to a Hagedorn transition. This statement of course needs to be verified on a more systematic level, but it is tempting to conjecture that this is a general situation, not limited to integrable theories and to low space-time dimensions. If so, would it mean that the majority of consistent $S$-matrices correspond to some kind of string theories? Or maybe there is a more general class of theories, besides the strings, which break the standard local structure of QFT while preserving macro-causality and exhibit the stringy density of states?

The present work represents a first step of a project having as a goal the systematic analysis of the TBA equations for completely general CDD-deformed factorizable $S$-matrices, with arbitrarily complicated CDD factors \eqref{cddg}, possibly including the factors \eqref{phientire} with singular behavior at high energies. 
Clearly, also CDD deformations of more complicated $S$-matrices, involving more than one kind of particles -- possibly having mass degeneracies, a situation leading to off-diagonal scattering -- have to be studied. Such $S$-matrices lead to systems of TBA equations more complicated than the simple equation \eqref{tbas}. Nonetheless, we believe that the numerical methods adopted here, in particular the PALC routine, can be adopted in full generality. Finally, a similar analysis can be extended to the CDD deformed ``massless TBA systems'' (see e.g. \cite{Zamolodchikov:1991vx,Zamolodchikov:1992zr,Fendley:1993jh}). Although the physical foundation here is less firm -- since the notion of $S$-matrix is ambiguous for massless theories in 1+1 dimensions -- these cases might yield welcome surprises.

\subsection*{Acknowledgements}
AZ acknowledges warm hospitality extended to him at the International Institute of Physics, Natal, Brasil, where parts of this work were done. AZ is grateful to A. Polyakov and F. Smirnov for interest and discussions.
SN wishes to thank R. Tateo, L. G. C\'{o}rdova and F. I. Schaposnik for their always interesting and useful comments and questions.  Work of GC was supported by MEC and MCTIC. Work of TF was supported by the Serrapilheira Institute (grant number Serra-1812-26900). ML was supported by the National Research Development and Innovation Office of Hungary under the postdoctoral grant PD-19 No. 132118 and by the Fund TKP2020 IES (Grant No. BME-IE-NAT), under the
auspices of the Ministry for Innovation and Technology. Early stage of ML's work was carried out at the International Institute of Physics, Natal, Brasil where he was supported by MEC and MCTIC.
Work of SN is supported by NSF grant PHY-1915219. Work of AZ was partly supported by the NSF grant PHY-191509.

\appendix
\section{Predictor-corrector routine}\label{app:pred_corr}
In general, a predictor-corrector routine is, as the name suggests, a two-step procedure to solve an equation, by first performing an educated (numerical) guess and subsequently adjusting it. In the case we are concerned with, we wish to solve the equation
\begin{eqnarray}
	H(\epsilon,R) = -\epsilon(\theta) + R \cosh\theta - \intop\,\frac{d\theta'}{2\pi} \varphi(\theta - \theta') \log\Big(1+e^{-\epsilon(\theta')}\Big) = 0\;,
\end{eqnarray}
with $\varphi$ being the 2CDD kernel \eqref{eq:2cdd_kernel}
\begin{eqnarray}
	\varphi(\theta) = \sum_{\sigma,\sigma' = \pm1}\frac{1}{\cosh(\theta + \sigma \omega + i \sigma'\gamma)}\;.
\end{eqnarray}
Obviously, we are going to deal with an appropriate truncation and discretization of the above equation, taking the following form
\begin{eqnarray}
	H_k(\vec{\epsilon},R) = -\epsilon_k + R \cosh\theta_k - \frac{1}{2\pi}\sum_{l}\Delta\theta \varphi_{kl}\log\Big(1+e^{-\epsilon_l}\Big) = 0\;,
\end{eqnarray}
with $\Delta\theta$ being the lattice step (taken to be constant, for simplicity) and
\begin{eqnarray}
	\varphi_{kl} = \sum_{\sigma,\sigma' = \pm1}\frac{1}{\cosh((k -l)\Delta\theta + \sigma \omega + i \sigma'\gamma)}\;.
\end{eqnarray}
The two steps of the predictor-corrector routine can be then described as follows
\begin{itemize}
	\item \textbf{Predictor}. This part of the routine takes as input a point $c(s_j) = (\vec{\epsilon}_j,R_j)$ on the solution curve and uses the initial value problem form (\ref{eq:in_val_prob_map_H}), which we recall here
	\begin{eqnarray}
		H'(c(s))\dot{c}(s) = \vec{0}\;,\qquad \vert\vert\dot{c}(s)\vert\vert = 1\;,\qquad c(s_j) = (\vec{\epsilon}_j,R_j)\;,
	\end{eqnarray}
	to yield a reasonable guess for a new point $c^{(0)}(s_{j+1}) = (\vec{\epsilon}^{\,(0)}_{j+1},R^{(0)}_{j+1})$. The simplest way to obtain such a point is to employ the so-called \emph{Euler predictor}, which implements the equation
	\begin{eqnarray}
		(\vec{\epsilon}^{\,(0)}_{j+1},R^{(0)}_{j+1}) = (\vec{\epsilon}_j,R_j) + \delta s\,\frac{t_j}{\vert\vert t_j\vert\vert}\;,
	\end{eqnarray}
	where the $N+1$ vector $t_j$ is tangent to the extended Jacobian $H'(c(s))$ at the point $ (\vec{\epsilon}_j,R_j) $:
	\begin{eqnarray}
		H' (\vec{\epsilon}_j,R_j) t_j = 0\;.
	\end{eqnarray}
	\item \textbf{Corrector}. This second part of the routine engages in the problem of adjusting the predictor's output $(\vec{\epsilon}^{\,(0)}_{j+1},R^{(0)}_{j+1})$ to a point actually lying on the solution curve. It does so by some iterative method for solving the equation $\vec{H} = 0$ starting from an initial, reasonably close, guess. The fastest and least expensive of these methods is the Newton's one, which in our case would take the following form
	\begin{eqnarray}
		\vec{\epsilon}^{\,(\ell+1)}_{j+1} = \vec{\epsilon}^{\,(\ell)}_{j+1} - [\mathcal{J}(\vec{\epsilon}^{\,(\ell)}_{j+1},R^{(\ell )}_{j+1})]^{-1}\, H(\vec{\epsilon}^{\,(\ell)}_{j+1},R^{(\ell )}_{j+1})\;,\qquad R^{(\ell +1)}_{j+1} = R^{(\ell +1)}_j
	\end{eqnarray}
	if only we were not worried to encounter a point where $\mathcal J$ is not invertible. In fact we are concerned precisely with such an eventuality, it being the very reason that led us to consider the PALC method and the associated predictor-corrector routine. Hence, we need to appropriately modify Newton's method in order to accommodate the possibility of a singular $\mathcal J$, with $H'$ of maximal rank $N$. The way to handle such a situation is to consider the concept of \emph{quasi-inverse} (also called \emph{Moore-Penrose inverse}) $A^{+}$ of a matrix $A$, defined as
	\begin{eqnarray}
		A^{+} = A^{T}\,(A\,A^{T})^{-1}\;,
	\end{eqnarray}
	where a superscript $T$ denotes standard matrix transposition. Notice that, if $A$ is a square matrix, the above definition is equivalent to the standard inverse. Now, if $A$ is instead an $N\times(N+1)$ matrix of maximal rank $N$ and $t$ is its tangent vector $At=0$, then the following statements are equivalent
	\begin{enumerate}
		\item $Ax = b$ \underline{and} $t^T x = 0$,
		\item $x = A^{+}b$,
		\item $x = \underset{v}{\textrm{min}}\Big[\,\vert\vert v\vert\vert \;\Big\vert\; Av = b \Big]$ which, in plain words, means that $x$ is the vector of minimal norm which solves the equation $Ax=b$.
	\end{enumerate}
	Without going too much in the details (see chapter 3 of \cite{allgower2012numerical}), the takeaway is that we can implement Newton's method in the usual way, as long as we trade the inverse of the Jacobian for the quasi-inverse of the extended Jacobian:
	\begin{eqnarray}
		(\vec{\epsilon}^{\,(\ell+1)}_{j+1},R^{(\ell +1)}_{j+1}) = (\vec{\epsilon}^{\,(\ell)}_{j+1},R^{(\ell )}_{j+1}) - [H'(\vec{\epsilon}^{\,(\ell)}_{j+1},R^{(\ell )}_{j+1})]^{+}\, H(\vec{\epsilon}^{\,(\ell)}_{j+1},R^{(\ell )}_{j+1})\;.
	\end{eqnarray}
	The above equation is then iterated as long as necessary, until reaching a point $(\vec{\epsilon}^{\,(L)}_{j+1},R^{(L)}_{j+1}) \equiv (\vec{\epsilon}_{j+1},R_{j+1})$ deemed, by some appropriate convergence test, close enough to a point on the solution curve.
\end{itemize}
Here follows a pseudo-code summarizing the procedure expounded above. As we can immediately see, the algorithm requires an initial point solving the TBA equation. This can be provided by using the standard iterative procedure of \S\ref{subsec:iterative} to solve the equation at some value of $R>R_*$. This will yield a solution $(\vec{\epsilon}_0,R_0)$ on the first branch, from which to start the PALC.
\begin{algorithm}[t!]
  \caption{Euler-Newton predictor-corrector routine}
\raggedright
  \textbf{Part 1: Input}\;
  \nl $(\vec{\epsilon}_0, R_0)$, s.t. $\vec{H}(\vec{\epsilon}_0, R_0)=\vec{0}$\hspace{164pt}INITIAL POINT\;
  \nl $\delta s$\hspace{307pt}STEP SIZE\;
  \nl $N_{\textrm{step}}$\hspace{264pt} STEP NUMBER\;
  \nl $\eta\ll1$\hspace{203pt}NUMERICAL TOLERANCE\;
  \textbf{Part 2: Initialization}\;
  \nl Solve $(\mathcal J)_0\,\vec{x} = -\frac{d}{d\lambda}\vec{H}_0$\hspace{133pt}FIND INITIAL TANGENT\;
  \nl $(\vec{t},\tau) = \frac{(\vec{x},1)}{\sqrt{1+\vert\vert\vec{x}\vert\vert^2}}$\hspace{168pt}NORMALIZE TANGENT\;
   \For{$j = 1$ \textbf{to} $N_{\textrm{step}}$}{
    \textbf{Part 3: Predictor}\;
    \nl Solve $\left(\begin{array}{c c} \mathcal J & \frac{d}{d\lambda}\vec{H} \\ \vec{t} & \tau\end{array}\right)_{j-1} \left(\begin{array}{c}\vec{t} \\ \tau\end{array}\right)_j = \left(\begin{array}{c} 0 \\ 1 \end{array}\right)$\hspace{44pt}FIND NEW TANGENT\;
    \nl $(\vec{\epsilon}_{j+1}^{(0)}, R_{j+1}^{(0)}) = (\vec{\epsilon}_{j}, R_{j})+ \delta s\frac{(\vec{t}_j,\tau_j)}{\vert\vert(\vec{t}_j,\tau_j)\vert\vert}$\hspace{83pt}EULER PREDICTOR\;
    \textbf{Part 4: Corrector}\;
    \For{$\ell = 0$ \textbf{to} $\infty\,,$ \textbf{until break}}{
      \nl $(\delta\vec{\epsilon},\delta R) = - \left[H'\left(\vec{\epsilon}_{j+1}^{(\ell)}, R_{j+1}^{(\ell)}\right)\right]^{+} \vec{H}\left(\vec{\epsilon}_{j+1}^{(\ell)}, R_{j+1}^{(\ell)}\right)$ \hspace{-6.5pt} CORRECTION STEP\;
      \nl $\left(\vec{\epsilon}_{j+1}^{(\ell+1)}, R_{j+1}^{(\ell+1)}\right) = \left(\vec{\epsilon}_{j+1}^{(\ell)}, R_{j+1}^{(\ell)}\right) + (\delta\vec{\epsilon},\delta R)$ \hspace{62.5pt}RELAXATION\;
      \nl \textbf{if} $\vert\vert\delta\vert\vert < \eta$ \textbf{BREAK}\hspace{84pt}CONVERGENCE CONDITION\;
    }
   \nl $\left(\vec{\epsilon}_{j+1}, R_{j+1}\right) = \left(\vec{\epsilon}_{j+1}^{(\ell)}, R_{j+1}^{(\ell)}\right)$\hspace{93pt} MOVE TO NEXT POINT\;
 }
\end{algorithm}
\newpage

\bibliography{biblio}

\end{document}